\documentclass[journal]{IEEEtran}

\usepackage[acronym]{glossaries}
\newacronym{BS}{BS}{base station}
\newacronym{LPF}{LPF}{low pass filter}
\newacronym{LO}{LO}{local oscillator}
\newacronym{PS}{PS}{phase-shifter}
\newacronym{PA}{PA}{power amplifier}
\newacronym{DDPG}{DDPG}{deep deterministic policy gradient}
\newacronym{STE}{STE}{straight through estimator}
\newacronym{RL}{RL}{reinforcement learning}
\newacronym{AP}{AP}{analog precoder}
\newacronym{LPA}{LPA}{linear power amplifier}
\newacronym{IL}{IL}{insertion loss}
\newacronym{CL}{CL}{convolutional layer}
\newacronym{FC-HBF}{FC-HBF}{fully-connected HBF}
\newacronym{FSA-HBF}{FSA-HBF}{fixed subarray HBF}
\newacronym{SA-HBF}{SA-HBF}{subarray HBF}
\newacronym{DSA-HBF}{DSA-HBF}{dynamic subarray HBF}
\newacronym{DUSA-HBF}{DUSA-HBF}{dynamic unequally subarray HBF}
\newacronym{BF}{BF}{beamforming}
\newacronym{UE}{UE}{user equipment}
\newacronym{AWGN}{AWGN}{additive white gaussian noise}
\newacronym{MIMO}{MIMO}{multiple-input multiple-output}
\newacronym{MISO}{MISO}{multiple-input single-output}
\newacronym{RF}{RF}{radio frequency}
\newacronym{RIS}{RIS}{reconfigurable intelligent surfaces}
\newacronym{IoT}{IoT}{internet-of-things}
\newacronym{ConvL}{ConvL}{convolutional layer}
\newacronym{FDD}{FDD}{frequency division duplex}
\newacronym{TDD}{TDD}{time-division duplex}
\newacronym{CSI}{CSI}{channel state information}
\newacronym{DNN}{DNN}{deep neural network}
\newacronym{DP}{DP}{digital precoder}
\newacronym{DL}{DL}{deep learning}
\newacronym{SVD}{SVD}{singular-value decomposition}
\newacronym{CNN}{CNN}{convolution neural network}
\newacronym{FDP}{FDP}{fully digital precoder}
\newacronym{SE}{SE}{spectral efficiency}
\newacronym{OFDM}{OFDM}{orthogonal frequency division multiplexing}
\newacronym{OMP}{OMP}{orthogonal matching pursuit}
\newacronym{FL}{FL}{fully-connected layer}
\newacronym{HSHO}{HSHO}{hybrid structured heuristic optimization}
\newacronym{DC}{DC}{direct current}
\newacronym{HBF}{HBF}{hybrid beamforming}
\newacronym{IA}{IA}{initial access}
\newacronym{mm-Wave}{mm-Wave}{millimeter wave}
\newacronym{mMIMO}{mMIMO}{massive MIMO}
\newacronym{SINR}{SINR}{signal-to-interference plus noise ratio}
\newacronym{SNR}{SNR}{signal-to-noise ratio}
\newacronym{SS}{SS}{synchronization signal}
\newacronym{SSB}{SSB}{synchronization signal burst}
\newacronym{RSSI}{RSSI}{received signal strength indicator}
\newacronym{PZF}{PZF}{phase zero forcing}
\newacronym{PSO}{PSO}{particle swarm optimization}
\newacronym{DAC}{DAC}{digital-to-analog converter}
\newacronym{ZF}{ZF}{zero forcing}
\newacronym{O-FDP}{O-FDP}{optimal fully digital precoder}
\newacronym{JT}{JT}{joint transmission}
\newacronym{CU}{CU}{central unit}
\newacronym{EE}{EE}{energy efficiency}
\newacronym{EC}{EC}{energy consumption}
\newacronym{MSE}{MSE}{mean square error}
\newacronym{CEL}{CEL}{cross entropy loss}
\newacronym{CB}{CB}{conjugate beamforming}
\newacronym{NC}{NC}{network controller}
\newacronym{CoMP}{CoMP}{coordinated multi point}
\newacronym{CF-mMIMO}{CF-mMIMO}{cell-free massive MIMO}
\newacronym{CF-HBF}{CF-HBF}{cell-free hybrid beamforming}
\newacronym{CF-BF}{CF-BF}{cell-free beamforming}

\newif\ifDeepMIMOModel
\DeepMIMOModeltrue

\newif\ifSimpleNParamEq
\SimpleNParamEqtrue

\usepackage{lettrine}

\usepackage{ifpdf}

\ifCLASSINFOpdf
  \usepackage[pdftex]{graphicx}
  \usepackage{graphicx,epstopdf}
  \graphicspath{{../pdf/}{../jpeg/}}
  \DeclareGraphicsExtensions{.pdf,.jpeg,.png,.eps}
\else
  \usepackage[dvips]{graphicx}
  \usepackage{graphicx,epstopdf}
  \graphicspath{{../eps/}}
  \DeclareGraphicsExtensions{}
\fi

\usepackage{cite}
\usepackage{algpseudocode}
\usepackage{amsthm}
\usepackage{balance}
\usepackage{eqparbox}
\usepackage{multirow}
\usepackage{amssymb}
\usepackage{array}
\usepackage{makecell}

\usepackage[ruled,vlined,noend]{algorithm2e}
\SetAlFnt{\small}

\usepackage{float}

\usepackage{amsmath}

\usepackage{mathtools, nccmath}

\usepackage{booktabs, siunitx}
\usepackage[svgnames,table]{xcolor}
\usepackage[scaled]{DejaVuSansMono}
\usepackage[T1]{fontenc}

\usepackage{color}
\usepackage{relsize}
\usepackage{mathtools}

\newcommand{\bs}[1]{\boldsymbol{#1}}

\newcommand{\mb}[1]{\mathbf{#1}}

\DeclareMathOperator*{\minimize}{minimize}

\SetKwInput{KwInput}{Input}                
\SetKwInput{KwOutputr}{Output Regression}              
\SetKwInput{KwOutputc}{Output Classification}              
\SetKwInput{KwOutput}{Output}              

\newcommand{\bseq}{\begin{subequations}}
\newcommand{\eseq}{\end{subequations}}
\newcommand{\baln}{\begin{align}}
\newcommand{\ealn}{\end{align}}
\newcommand{\balnd}{\begin{aligned}}
\newcommand{\ealnd}{\end{aligned}}
\newcommand{\beq}{\begin{equation}}
\newcommand{\eeq}{\end{equation}}
\newcommand{\beqn}{\begin{eqnarray}}
\newcommand{\eeqn}{\end{eqnarray}}
\newcommand{\beqno}{\begin{eqnarray*}}
\newcommand{\eeqno}{\end{eqnarray*}}
\newcommand{\bma}{\begin{displaymath}}
\newcommand{\ema}{\end{displaymath}}
\newcommand{\bnu}{\begin{enumerate}}
\newcommand{\enu}{\end{enumerate}}
\newcommand{\bce}{\begin{center}}
\newcommand{\ece}{\end{center}}
\newcommand{\btb}{\begin{tabular}}
\newcommand{\etb}{\end{tabular}}
\newcommand{\ba}{\begin{array}}
\newcommand{\ea}{\end{array}}
\usepackage{footnote}
\makesavenoteenv{tabular}
\makesavenoteenv{table}

\usepackage{algpseudocode}

\usepackage{tikz}
\usetikzlibrary{decorations.pathreplacing,calc}
\newcommand{\tikzmark}[1]{\tikz[overlay,remember picture] \node (#1) {};}

\newcommand*{\SpaceReservedForComments}{1cm}%
\newcommand*{\HorizontalOffset}{-0.5em}%
\newcommand*{\VerticalOffset}{0.7ex}%
\newcommand*{\AddNote}[4][]{%
    \begin{tikzpicture}[overlay, remember picture]
        \draw [decoration={brace,amplitude=0.5em},decorate,ultra thick,red, #1]
            ($(#3)+(\HorizontalOffset,-\VerticalOffset)$) --  ($(#2)+(\HorizontalOffset,\VerticalOffset)$)
            node [align=left, text width=\SpaceReservedForComments+1.0em, pos=0.5, anchor=east] {#4};
    \end{tikzpicture}
}%

\makeatletter
    \algrenewcommand\alglinenumber[1]{\tikzmark{\arabic{ALG@line}}\tiny#1:}
\makeatother

\begin{document}
\title{Learning Energy-Efficient Hardware Configurations for Massive MIMO Beamforming}

\author{\IEEEauthorblockN{Hamed Hojatian,~\IEEEmembership{Member,~IEEE,}
Zoubeir Mlika,~\IEEEmembership{Member,~IEEE,} J\'{e}r\'{e}my Nadal,~\IEEEmembership{Member,~IEEE,}\\
Jean-Fran\c{c}ois Frigon,~\IEEEmembership{Senior Member,~IEEE,} 
 and Fran\c{c}ois Leduc-Primeau,~\IEEEmembership{Member,~IEEE}}
\thanks{This work was supported by the Natural Sciences and Engineering Research Council of Canada (NSERC) (under project ALLRP 566589-21) and InnovÉÉ (INNOV-R program) through the partnership with Ericsson and ECCC.
H. Hojatian, Zoubeir Mlika, J-F. Frigon, F. Leduc-Primeau are with the Department of Electrical Engineering, Polytechnique Montreal, Montreal, QC, H3T lJ4, Canada.  J. Nadal is with the Department of Mathematical and Electrical Engineering, IMT Atlantique, France. Corresponding author: H. Hojatian (Emails: hamed.hojatian@polymtl.ca)}}


\IEEEtitleabstractindextext{%
\begin{abstract}
Hybrid beamforming (HBF) and antenna selection are promising techniques for improving the energy efficiency~(EE) of massive multiple-input multiple-output~(mMIMO) systems. However, the transmitter architecture may contain several parameters that need to be optimized, such as the power allocated to the antennas and the connections between the antennas and the radio frequency chains. Therefore, finding the optimal transmitter architecture requires solving a non-convex mixed integer problem in a large search space. In this paper, we consider the problem of maximizing the EE of fully digital precoder~(FDP) and hybrid beamforming~(HBF) transmitters. 
First, we propose an energy model for different beamforming structures. 
Then, based on the proposed energy model, we develop an unsupervised deep learning method to maximize the EE by designing the transmitter configuration for FDP and HBF. The proposed deep neural networks can provide different trade-offs between spectral efficiency and energy consumption while adapting to different numbers of active users. 
Finally, to ensure that the proposed method can be implemented in practice, we investigate the ability of the model to be trained exclusively using imperfect channel state information~(CSI), both for the input to the deep learning model and for the calculation of the loss function. Simulation results show that the proposed solutions can outperform conventional methods in terms of EE while being trained with imperfect CSI. Furthermore, we show that the proposed solutions are less complex and more robust to noise than conventional methods.
\end{abstract}

\begin{IEEEkeywords}
Hybrid Beamforming, Subarray Hybrid Beamforming, Fully Digital Beamforming, Deep Neural Network, Unsupervised Learning, Energy Efficiency. 
\end{IEEEkeywords}}

\maketitle
\IEEEdisplaynontitleabstractindextext

\IEEEpeerreviewmaketitle

\section{Introduction} \label{sec:intro}

Wireless communication has been revolutionized by \gls{mMIMO} technologies, which are already one of the key enabling technologies in the fifth-generation~(5G) of wireless networks thanks to their potential to increase the transmission capacity through the deployment of large-scale antenna arrays at the transmitter or receiver side~\cite{Marzetta2015}.
As a result, \gls{mm-Wave} communications can be used at longer ranges, thus greatly increasing the bandwidth available to wireless networks \cite{mm-wave1}.

The conventional implementation of MIMO systems utilizes a dedicated \gls{RF} chain for each antenna element. Even though this approach is appropriate for common small-scale MIMO systems, it is inadvisable for \gls{mMIMO} systems equipped with a large number of antenna elements due to the high production costs and power consumption associated with the RF circuitry. Therefore, even though \gls{mMIMO} is an important technology for future generations of wireless networks, it still faces many technical challenges to improve its \gls{EE} and, to date, it remains a subject of ongoing research~\cite{9064949}. In light of this, \gls{HBF} and antenna selection are proposed as an effective way to facilitate the implementation and to improve the \gls{EE} of \gls{mMIMO} systems~\cite{hbf_survey}. Indeed HBF reduces the number of RF chains and \glspl{DAC}, helping to improve \gls{EE}. Accordingly, for better \gls{EE}, \gls{HBF} techniques are being examined for 5G cellular networks in the \gls{mm-Wave} frequency bands, and will likely also be found in sixth-generation (6G) networks~\cite{8808168}.

Different \gls{HBF} structures have been proposed to achieve different trade-offs between cost, energy consumption, and \gls{SE}, which can be grouped into three general categories, \gls{FC-HBF}~\cite{widebandchhbf}, \gls{FSA-HBF}~\cite{7436794}, and \gls{DSA-HBF}~\cite{7880698}. Each category has its advantages and limitations. \gls{FC-HBF}  offers flexibility but has higher implementation complexity. \gls{FSA-HBF} balances \gls{SE} and complexity. Finally, \gls{DSA-HBF} provides adaptability but with additional design complexities. To configure a \gls{HBF} structure, one of the most prominent techniques consists in minimizing the Euclidean distance between the desired \gls{FDP} and its hybrid counterpart~\cite{yu_tsp_16}. However, this technique requires designing the \gls{FDP}, which is computationally complex and not necessarily energy-efficient. Furthermore, the number of possible \gls{HBF} structures is extremely large, making it complicated to find an optimal HBF structure. Therefore, the question that arises is how to efficiently design the best HBF structure in terms of energy consumption and \gls{SE}. Towards answering this question, our first step consists of proposing an accurate energy model that finds the power consumption of each component in different beamforming structures. Our second step involves applying machine learning-based approaches to design the beamforming structure instead of using complex optimization-based ones.

Thanks to the enormous success of \gls{DL}, in a wide variety of engineering fields, \glspl{DNN} have received significant attention in recent years and have been widely applied to wireless communication systems~\cite{10002368, 8382166}. Despite the fact that training \glspl{DNN} to solve wireless communication problems can be computationally intensive, it can take place offline and only the trained DNN model will be used to make online decisions, thus reducing the overall complexity. Different studies used \glspl{DNN} to address complex problems within the physical layer~\cite{8663966}. In supervised learning approaches, the time spent during the data labeling procedure is not negligible. In addition, this procedure must be performed each time a new dataset is used for training. In reinforcement learning (RL) approaches, an agent collects online data as it interacts with its environment in a trial-and-error manner. In \gls{mMIMO} systems, since the \gls{HBF} action space is large, the convergence of the RL model requires a large number of experiments. As a consequence, unsupervised learning demonstrates superiority over supervised learning and reinforcement learning in terms of its ability to autonomously extract meaningful patterns and insights from large datasets without relying on explicit labels or large training overhead.

To summarize, in this study, we aim to optimize the \gls{EE} of mmWave \gls{mMIMO} systems by designing \gls{HBF} structures and \gls{FDP} using DL-based techniques. The problem consists of jointly designing the transmitter configuration and beamforming weights that maximize the \gls{EE}. To accomplish this, we first propose an accurate energy model that takes into account the power consumption of the different components of the \gls{mMIMO} system. Second, we propose an unsupervised deep learning approach that incorporates two key components to design an energy-efficient beamforming structure: (i) a novel loss function that considers different trade-offs between \gls{SE}, energy consumption, and active users, and (ii) imperfect \gls{CSI} during both the training and inference phases.


\subsection{Related Works}
In \cite{7370753}, the authors compared the EE of six different \gls{PS}-based and switch-based \gls{HBF} structures. 
However, given the hardware available today, the energy model in \cite{7370753} overstates the power consumption of \glspl{PS}, which makes the conclusion unfair to \gls{PS}-based approaches. 
Many studies are proposed in the context of \gls{DL}-aided \gls{HBF} design and antenna selection algorithms~\cite{hojatian2020rssi, HBF, 9189869, 9112250, nguyen2023deep, unsupervised, 9729183, 9681824,9337188,8924932,8446042,7523998,9456019}. In particular, a \gls{RSSI}-based \gls{FC-HBF} design implemented with supervised learning is proposed in~\cite{hojatian2020rssi}. The authors of~\cite{HBF} suggested a supervised learning approach for \gls{FC-HBF} design under perfect \gls{CSI}. Another form of supervised learning is also proposed for the \gls{FSA-HBF} design with perfect \gls{CSI} in~\cite{9189869}. The authors of~\cite{9112250} proposed a \gls{RL} approach to design the \gls{HBF}. However, they assumed that the \gls{CSI} is known perfectly, and due to the continuous action space, their method relies on deep deterministic policy gradient~(DDPG), which is computationally complex~\cite{DDPG_cons}. In the context of unsupervised learning for the \gls{FC-HBF} design, the authors in~\cite{unsupervised, 9729183} presented a novel \gls{HBF} design employing imperfect \gls{CSI} for single \gls{BS} and cell-free \gls{mMIMO} (CF-mMIMO), respectively. However, their approaches are only for \gls{FC-HBF}. In~\cite{9681824}, the authors proposed an unsupervised learning approach for \gls{HBF} and antenna selection using a differentiable activation function for 1-bit \glspl{PS}. However, the main objective is to maximize the \gls{SE} of the \gls{mMIMO} system, and the authors neither optimized the \gls{EE} nor considered an accurate energy model. In~\cite{9337188, 8924932, 8446042}, the authors proposed a joint antenna selection and precoding design with an iterative algorithm and a \gls{DL} solution to maximize the \gls{SE} of multi-user multiple-antenna downlink systems. The proposed ML approach assumes perfect \gls{CSI} for the training data, does not optimize the \gls{EE}, and requires a complex iterative algorithm to generate the training samples. In~\cite{7523998, 9456019}, the authors proposed a supervised learning approach to solve the antenna selection problem. However, the proposed method only applies to \gls{FDP}.

\subsection{Contributions}
In this paper, we consider both \gls{FDP} and \gls{HBF} transmitters and develop new unsupervised deep learning solutions that jointly design beamforming and antenna selection while taking into account the power consumption and \gls{IL} of all components. For \gls{FDP}, the proposed solution designs the \gls{FDP} vectors along with the antenna selection solution, while for \gls{HBF}, thanks to a multi-tasking \gls{DNN}, the proposed solution directly provides the \gls{AP} and \gls{DP} with the power allocation among the antennas. A preliminary version of this work was published in~\cite{Flexible}, where we only considered maximizing the \gls{SE} by designing the \gls{HBF} for fixed and dynamic HBF structures.

In summary, the contributions of this work are as follows:
\begin{itemize}
    \item We propose an accurate energy model for the \gls{FDP} and \gls{HBF} structures while considering the latest state-of-the-art hardware solutions.
    \item We propose an unsupervised deep learning solution robust against imperfect \gls{CSI} to find the optimal energy-efficient antenna selection for \gls{FDP} and transmit power allocation for HBF considering the proposed accurate energy model. Due to the binary constraints of beamforming connections, our unsupervised deep learning approach makes use of the Gumbel-Sigmoid technique inspired by Gumbel-Softmax. The Gumbel-Sigmoid technique is designed such that it considers the constraints of all components involved in the beamforming connections.
    \item We design an unsupervised loss function that takes into account the \gls{SE}, the \gls{EC} as well as the number of active users. Thanks to this loss function, the proposed solution is flexible and can intelligently adjust the power consumption according to the number of active users and can provide an optimal trade-off between \gls{SE} and \gls{EC}.
    \item We train the proposed unsupervised deep learning solution using imperfect \gls{CSI} for both the \gls{DNN} input and the loss function computation. We also investigate the noise tolerance of our approach by showing that imperfect inputs can be beneficial and improve the \gls{EE} of the \gls{mMIMO} system.
    \item The proposed solutions are evaluated in a realistic ray-tracing channel model generated using a three-dimensional model of an urban environment to capture the geometry-based characteristics of the channel. The simulation results show that the proposed solution outperforms conventional solutions in terms of \gls{EE} with lower computational complexity, and can be adapted to achieve different trade-offs between \gls{SE} and \gls{EC}.
\end{itemize}

\subsection{Paper Organization and Notation}
The rest of the paper is organized as follows:
 In Section~\ref{sec:system}, we present the system setup followed by the baseline solutions and the channel model. The proposed energy model for the different beamforming structures is provided in Section~\ref{sec:Energ}.
 Section~\ref{sec:Optimal} presents the proposed energy-efficient unsupervised learning solutions for HBF and FDP, including a discussion of the DNN structure, the training phase, and the online phase. In Section~\ref{sec:sim}, we evaluate the performance of the proposed algorithms by comparing them with state-of-the-art solutions using a realistic ray-tracing channel model. Finally, Section~\ref{sec:Conc} concludes the paper.

Matrices, vectors, and scalars are denoted by boldface uppercase, boldface lowercase, and normal letters, respectively. The notations $(\cdot)^{\rm{H}}$, $(\cdot)^{\rm T}$, $(\cdot)^{\dagger}$, $|\cdot|$, $\|\cdot\|_{F}$, $\|\cdot\|_2$, $\Re[\cdot]$, $\Im[\cdot]$, $\mathbb{I}_{n}$, $\otimes$ denote Hermitian transpose, transpose, Moore-Penrose pseudoinverse, absolute value, Frobenius norm, $\ell^2$-norm, real part, imaginary part, the $n \times n$ identity matrix, and element-wise product, respectively.

\section{System Model and Baselines} \label{sec:system}

\begin{figure}[t!]
    \centering
    \includegraphics[width=0.9\columnwidth]{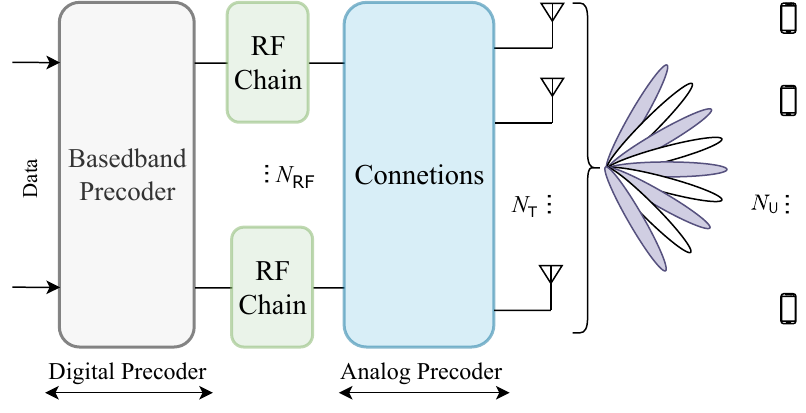}
    \caption{Massive MIMO system model structure with one transmitter BS employing HBF to serve a set of users.}
    \label{fig:HBF_arch}
\end{figure}

Let us assume a \gls{TDD} multi-user mMIMO system where channel reciprocity is available such that the uplink channel estimate can be used for the downlink transmission. The \gls{mMIMO} system consists of a single \gls{BS} in a single-cell equipped with $N_{\sf{T}}$ antennas and $N_{\sf{RF}}$ RF chains serving $N_{\sf{U}}$ single antenna users simultaneously as shown in Figure~\ref{fig:HBF_arch}. The \gls{DP} is performed in the baseband and then the output signal goes through the RF chains. Each RF chain is composed of a \gls{DAC}, a \gls{LPF}, a \gls{LO} and a mixer, and is connected to the $N_{\sf{T}}$ antennas. The resolution for all \glspl{DAC} and \glspl{PS} are fixed respectively to $b_{D}$ and $q$. The RF chains are connected to the antennas through \glspl{PS}. The network of these connections and \glspl{PS}, known as \gls{AP}, can be tuned based on different \gls{HBF} structures.

\subsection{Conventional Beamforming Structures} \label{Sec:Baseline}
We first review the three conventional beamforming structures followed by their non-DL design methods, which are used as baselines for our proposed solutions. Connections between the RF chains and the antennas are represented by a binary matrix $\bs{\Omega}$. For HBF, $\bs{\Omega}=\bs{\Omega}_{\sf{HB}}\in\{0,1\}^{N_{\sf{T}} \times N_{\sf{RF}}}$, and $[\bs{\Omega}_{\sf{HB}}]_{n,m}=1$ if antenna $n$ is connected to RF chain $m$. For FDP, $\bs{\Omega}=\bs{\Omega}_{\sf{FD}}$ is an $N_{\sf{T}} \times N_{\sf{T}}$ diagonal binary matrix, with $\bs{\Omega}_{\sf{FD}} = \text{diag}(\bs{\omega})$, where $\bs{\omega} = [\omega_{1}, \ldots, \omega_{N_{\sf{T}}}]$ and $\omega_{n}=1$ if antenna $n$ is activated.

\subsubsection{Fully Digital Precoder~(FDP)} \label{subsub:FDP}
In \gls{FDP}, each antenna is connected to an RF chain through the circuit of DAC, LPF, LO, and a mixer. The signal received by each user can be written as
\begin{equation} \label{recv_sig_FDP}
\mb{y}_u =  \mb{h}_{u}^{\rm{H}} \sum_{u=1}^{N_{\sf{U}}}  \mb{u}_{u} x_u + \bs{\eta},
\end{equation}
where $\mb{h}_{u} \in \mathbb{C}^{N_{\sf{T}} \times 1}$ stands for the channel vector from the $N_{\sf{T}}$ antennas of the \gls{BS} to the user index $u$, $\mb{x} = [x_1,\ldots, x_u,\ldots, x_{N_{\sf{U}}}]$ is the matrix of transmitted symbols for all users, normalized to $\mathop{\mathbb{E}}[\mb{x}\mb{x}^{\rm{H}}] = \frac{1}{N_{\sf{U}}} \mathbb{I} _{N_{\sf{U}}}$, $\bs{\eta}$ is the \gls{AWGN} with mean 0 and variance $\sigma^2$, and $\mb{u}_u$ denotes the precoder vector for the $u^{\text{th}}$ user. The \gls{SE} of the \gls{O-FDP}, $\mb{U}_{\sf{opt}}=[\mb{u}_{1},\ldots, \mb{u}_{u},\ldots,  \mb{u}_{N_{\sf{U}}}]$ for single-antenna users is obtained by solving the following problem:
\begin{subequations} \label{eq:FDP_opti_problem}
\begin{eqnarray} 
&\underset{\left\lbrace \mb{U}_{\sf{opt}} \right\rbrace}{\max} &
R_{\sf{FDP}}(\mb{U}_{\sf{opt}})  \\
& \text{ s. t.} & \sum_{u=1}^{N_{\sf{U}}} \mb{u}_{u}^{\rm{H}} \mb{u}_{u} \leq P_{\sf{TX}}\, , \label{cnt_op2}
\end{eqnarray}
\end{subequations}
where $R_{\sf{FDP}}(\mb{U}_{\sf{opt}}) = \sum_{u=1}^{N_{\sf{U}}} \log_2(1 + \text{SINR}(\mb{u}_u)) \,$, the \gls{SINR} of the $u^{\text{th}}$ user is given by
\begin{align}
    \label{eq:FDP_SINR}
    \text{SINR}(\mb{u}_{u}) = \frac{ \big|\mb{h}^{\rm{H}}_{u} \mb{u}_{u} \big|^2}{\sum\limits_{\substack{j=1\\j\neq u}}^{N_{\sf{U}}} \big|\mb{h}^{\rm{H}}_{u} \mb{u}_{j} \big|^2 + \sigma^2}\,,
\end{align}
and $P_{\sf{TX}}$ is the normalized total transmit power constraint.
The baseline results presented in this paper are obtained by solving \eqref{eq:FDP_opti_problem} based on~\cite{BBO2014}. We refer to this approach as optimal \gls{FDP}~(O-FDP). For a given connection matrix $\bs{\Omega}_{\sf{FD}}$, the FDP is given by $\mb{U} = \bs{\Omega}_{\sf{FD}} \times \mb{U}_{\sf{opt}}$.

\begin{figure}[t!]
    \centering
    \includegraphics[width=0.8\columnwidth]{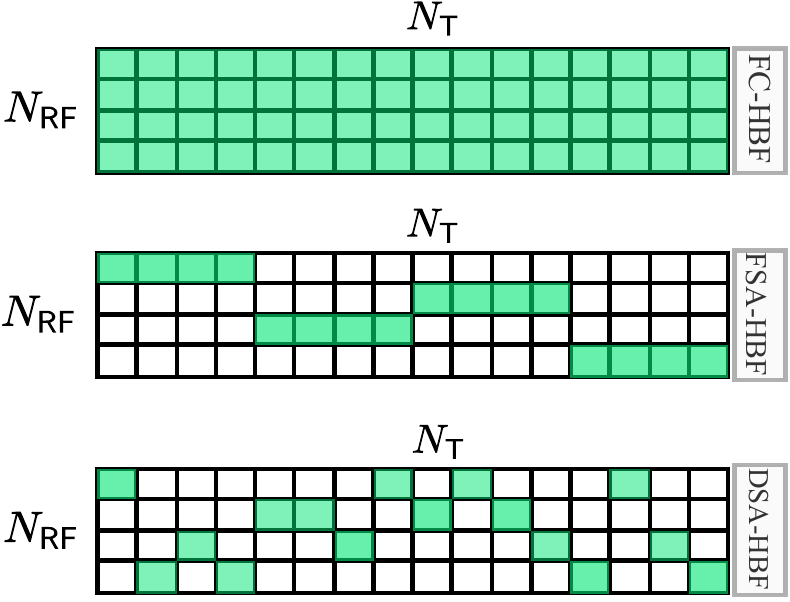}
    \caption{Example of connection matrices~($\bs{\Omega}_{\sf{HB}}$) of some conventional HBF structures. Each green square indicates a connection. Top: FC-HBF. Middle: FSA-HBF. Bottom: DSA-HBF.}
    \label{fig:HBF_types}
\end{figure}

\subsubsection{Fully Connected Hybrid Beamforming~(FC-HBF)} 

In all \gls{HBF} structures, we assume $N_{\sf{RF}} << N_{\sf{T}}$.
Regardless of the chosen \gls{HBF} structure, the signal received by each user can be written as
\begin{equation} \label{recv_sig_HBF}
\mb{y}_u =  \mb{h}_{u}^{\rm{H}} \mb{A} \sum_{u=1}^{N_{\sf{U}}}  \mb{w}_{u} x_u + \bs{\eta} \, .
\end{equation}
 The \gls{HBF} vectors consist of a \gls{DP}, $\mb{W} = [\mb{w}_{1},\ldots, \mb{w}_{u}, \ldots, \mb{w}_{N_{\sf{U}}}] \in \mathbb{C}^{N_{\sf{RF}} \times N_{\sf{U}}}$, and an \gls{AP}, $\mb{A} \in \mathbb{C}^{N_{\sf{T}} \times N_{\sf{RF}}}$. Since the AP is a combination of the \glspl{PS} and combiners and it depends on \gls{HBF} structure and connection between the antennas and RF chains, we define it as follows: \beq \label{Eq:analog}
\mb{A} = \mb{P}_{q} \otimes \bs{\Omega}_{\sf{HB}},
\eeq
where $\mb{P}_{q} \in \mathbb{C}^{N_{\sf{T}} \times N_{\sf{RF}}}$ is the coefficient of the $q$ bits \gls{PS} connecting the $n^{\text{th}}$ antenna and $m^{\text{th}}$ RF chain, where $[\mb{P}_q]_{n,m} \in \{e^{j2\pi k/2^{q}} : k\in\{1,\dots,2^{q}\} \}$. 
Therefore, the \gls{SE} for a given HBF $(\mb{A},\mb{W})$ is given by
\begin{align}
\label{eq:sumRate_HBF} 
    R_{\sf{HBF}}(\mb{A}, \mb{W} ) = \sum_{u=1}^{N_{\sf{U}}}\text{log}_2 \bigl(  1+ \text{SINR}(\mb{A}, \mb{w}_{u}) \bigr) \, ,
\end{align}
and the \gls{SINR} of the $u^{\text{th}}$ user can be expressed as
\begin{align}
    \label{eq:SINR_HBF}
    \text{SINR}(\mb{A}, \mb{w}_{u}) & = \frac{ \big|\mb{h}^{\rm{H}}_{u} \mb{A} \mb{w}_{u} \big|^2}{\sum\limits_{\substack{j=1\\j \neq u}}^{N_{\sf{U}}} \big|\mb{h}^{\rm{H}}_{u}\mb{A} \mb{w}_{j} \big|^2 + \sigma^2} \, .
\end{align}

In \gls{FC-HBF}, we set $\bs{\Omega}_{\sf{HB}} = \bs{\Omega}_{\sf{FC}}$, and all RF chains are connected to all antennas through \glspl{PS}, combiners, and \gls{PA} as shown in Figure~\ref{fig:HBF_types}~(top), where the green boxes show the connections. This structure thereby enables maximum design flexibility and therefore requires a large number of \glspl{PS} and combiners, which increase the implementation cost and energy consumption.
The AP of the FC-HBF can be expressed according to~\eqref{Eq:analog} with
\beq
\label{eq6}
[\bs{\Omega}_{\sf{HB}}]_{n,m} = 1 \quad \forall n,m \, ,
\eeq
\beq
\label{eq6a}
[\mb{P}_q]_{n,m} \in \{e^{j2\pi k/2^{q}} : k\in\{1,\dots,2^{q}\} \} \quad \forall n,m \, .
\eeq
Since all the antennas are connected to all the RF chains through a \gls{PS} with $q$-bit quantization, the feasible analog precoder for $n^{\text{th}}$ antenna and $m^{\text{th}}$ RF chain is $[\mb{A}]_{n,m} \in \{e^{j2\pi k/2^{q}} : k\in\{1,\dots,2^{q}\} \}$.
Conventional \gls{HBF} solutions either rely on codebook-based solutions to limit the number of feasible solutions \cite{ayach2014} or, more rarely, use real-valued \glspl{PS} \cite{yu_tsp_16}. The conventional approach consists of first designing the \gls{O-FDP} matrix in \eqref{eq:FDP_opti_problem}. Then, the \gls{AP} and \gls{DP} are designed in such a way that the resulting precoders approximate $\mb{U}_{\sf{opt}}$ as follows:
\begin{subequations} \label{eq:MMSE_prb}
\begin{eqnarray} 
&\underset{\mb{A}, \mb{W}} {\minimize\;} & \big\|\mb{U}_{\sf{opt}} - \mb{A} \mb{W} \big\|^2_{F} \\
& \text{ s. t.} & \eqref{Eq:analog},\eqref{eq6},\eqref{eq6a},\|\mb{A}\mb{W}\|_F^2=N_{\sf{U}}.\label{cnt_2_MMSE}
\end{eqnarray}
\end{subequations}

We obtain the \gls{FC-HBF} solution of~\eqref{eq:MMSE_prb} using ``PE-AltMin'' and ``MO-AltMin'' proposed in~\cite{yu_tsp_16}. 

\subsubsection{Subarray Hybrid Beamforming} \label{Sec:FSA}
Each antenna in a subarray structure is connected to only one RF chain through a \gls{PS}. Consequently, the total number of \glspl{PS} is reduced to $N_{\sf{T}}$, instead of $N_{\sf{T}}\times N_{\sf{RF}}$ in the FC-HBF. In the subarray HBF structure, we consider two types of connection: (i) a structure equipped with fixed connections, known as fixed subarray HBF (FSA-HBF), or (ii) a structure equipped with dynamic connections, known as dynamic subarray HBF (DSA-HBF). Examples of possible connection matrices for each case are shown in Figure~\ref{fig:HBF_types}.
The DSA-HBF structure enables the antennas and the RF chains to be dynamically switched at each time interval in response to changing conditions. It was shown that such a dynamic structure significantly enhances the \gls{SE} of the system by providing more degrees of freedom in the \gls{HBF} design compared to a \gls{FSA-HBF} structure, and reduces the power consumption compared to the FC-HBF structure \cite{7880698}. Therefore, based on the general definition of the \gls{AP}~($\mb{A} = \mb{P}_{q} \otimes \bs{\Omega}_{\sf{HB}}$), the constraint on matrix $\bs{\Omega}_{\sf{HB}}$ for subarray \gls{HBF} is given by
\begin{equation}\label{DSA_cons}
      \sum_{m=1}^{N_{\sf{RF}}} [\bs{\Omega}_{\sf{HB}}]_{n,m} = 1 \quad \forall \ n \, .
\end{equation}
To find the precoder matrices for \gls{FSA-HBF}, the general approach described in \eqref{eq:MMSE_prb} for \gls{FC-HBF} can be used.
For the \gls{DSA-HBF}, the connection pattern ($\bs{\Omega}_{\sf{HB}}$) between the RF chains and the antennas is dynamic and needs to be optimized, resulting in a large design space.

\subsection{Problem Definition} \label{subsec:problem}
The main objective of this paper is to maximize the EE of the mMIMO system by  selecting the antennas and designing the BF structure. 
For the \gls{FDP} case, the problem consists of finding the precoder matrix $\mb{U}$ and antenna selection $\bs{\Omega}_{\sf{FD}}=\text{diag}(\bs{\omega})$ that maximize the \gls{EE}, while achieving a desired minimum average \gls{SE} denoted as $R_{\sf{d}}$. More formally, we seek to solve the following optimization problem:
\begin{subequations} \label{eq:FDP_problem}
\begin{eqnarray} 
&\underset{\mb{U}, \bs{\Omega}_{\sf{FD}}}{\text{maximize}} &
R_{\sf{FDP}}(\mb{U} \times \bs{\Omega}_{\sf{FD}}) / P_{\text{FDP}} \\
& \text{ s. t.} & \sum_{u=1}^{N_{\sf{U}}} \mb{u}_{u}^{\rm{H}} \mb{u}_{u} \leq P_{\sf{TX}} \,, \\ \label{cnt_op2_FDP}
& & [\bs{\Omega}_{\sf{FD}}]_{n,n} \in \{0,1\}, \;\forall n, \\
& & R_{\sf{FDP}}(\mb{U} \times \bs{\Omega}_{\sf{FD}}) \geq N_{\sf{U}} R_{\sf{d}},
\end{eqnarray}
\end{subequations}
where $P_{\text{FDP}}$ is the total power consumed by the BF components. 

Similarly, the \gls{HBF} design consists in finding the precoder matrices $\mb{W}$ and $\mb{A}$ and the power allocation that maximizes the EE. Therefore, we have the following optimization problem:
\begin{subequations}  \label{SEE-max-prb}
\begin{eqnarray} 
&\underset{\mb{A}, \mb{W}}{\text{maximize}} &  R_{\sf{HBF}}(\mb{A}, \mb{W}) / P_{\text{HBF}}, \\
& \text{s.t.} & \sum_{u=1}^{N_{\sf{U}}} \mb{w}_{u}^{\rm{H}} \mb{A}^{\rm{H}} \mb{A} \mb{w}_{u} \leq P_{\sf{TX}}\, , \\ \label{cnt2}
&& \mb{A} = \mb{P}_{q} \otimes \bs{\Omega}_{\sf{HB}}, \\
&& [\bs{\Omega}_{\sf{HB}}]_{n,m} \in \{0,1\},\; \forall n ,m, \\
& & R_{\sf{HBF}}(\mb{A}, \mb{W}) \geq N_{\sf{U}} R_{\sf{d}},
\end{eqnarray}
\end{subequations}%
where $P_{\text{HBF}}$ is the total power consumed by the \gls{HBF} transmitter, and again $R_{\sf{d}}$ is the minimum average required \gls{SE}. The power consumption $P_{\text{FDP}}$ and $P_{\text{HBF}}$ will be described in detail in Section~\ref{sec:Energ}.
In this paper, for simplicity, we consider a total power constraint for the transmitter, where the power transmitted by each antenna is not necessarily equal or limited to $P_{\sf{TX}} / N_{\sf{T}}$. It should be noted that in HBF turning off an antenna is not necessarily corresponding to deactivating an RF chain. On the contrary, since in FDP, each antenna is connected to one RF chain, and the power consumed by RF chains is noticeable, turning off the RF chains leads to the deactivation of the corresponding antennas.

\subsection{Channel Model}\label{subsec:channel}
The experiments presented in this paper are based on
the generic deep learning dataset for \gls{mm-Wave} mMIMO systems (known as deepMIMO)~\cite{deepmimo}, which provides a channel vector $\mb{h}$ of length $N_{\sf{T}}$ for each user position on a quantized grid. An $N_{\sf{T}} \times N_{\sf{U}}$ channel matrix entries in the dataset are obtained by concatenating the $N_{\sf{U}}$ channel vectors randomly selected from the available user positions of the considered area. 

Since we consider \gls{TDD} communication with channel reciprocity, the estimated \gls{CSI} in the uplink can be employed for downlink. However, due to channel estimation errors, the downlink channel cannot be perfectly estimated. 
Thus, to model the channel estimation error, the \gls{BS} uses the minimum mean square error such that the estimated channel matrix is given by \cite{6375940}:
\beq \label{CSI_N}
\hat{\mb{H}} = \sqrt{1 - \beta^2} \mb{H} + \beta \bs{\epsilon},
\eeq
where $\mb{H} = [\mb{h}_{1}, ..., \mb{h}_{N_{\sf{U}}}]^{\rm{T}}$ is the actual channel matrix, the scaling coefficient $\beta \in [0,1]$ represents the reliability of the estimate, and $\bs{\epsilon}\sim\mathcal{N}(0,\sigma_e^2)$ is an error matrix modeled as a zero-mean Gaussian noise with variance $\sigma_e^2$. Unlike previous \gls{DL}-based studies, where perfect CSI is available during the training of the \gls{DNN}, in this work, we propose to use the imperfect CSI~($\hat{\mb{H}}$) not only as the input to the DNN but also to compute the loss function during the training phase. In Section~\ref{sec:sim}, we further evaluate the impact of the imperfect \gls{CSI} by varying the value of $\beta$ and show that a moderate level of imperfection in CSI can act as a regularizer for the \gls{DNN} and slightly improve the \gls{SE}.

\section{Energy Model} \label{sec:Energ}

In this section, we present an energy model for the different \gls{FDP} and \gls{HBF} hardware configurations, considering both the direct energy consumption as well as the energy consumption resulting from \gls{IL} of each component.

\begin{figure}[t]
    \centering
    \includegraphics[width=\columnwidth]{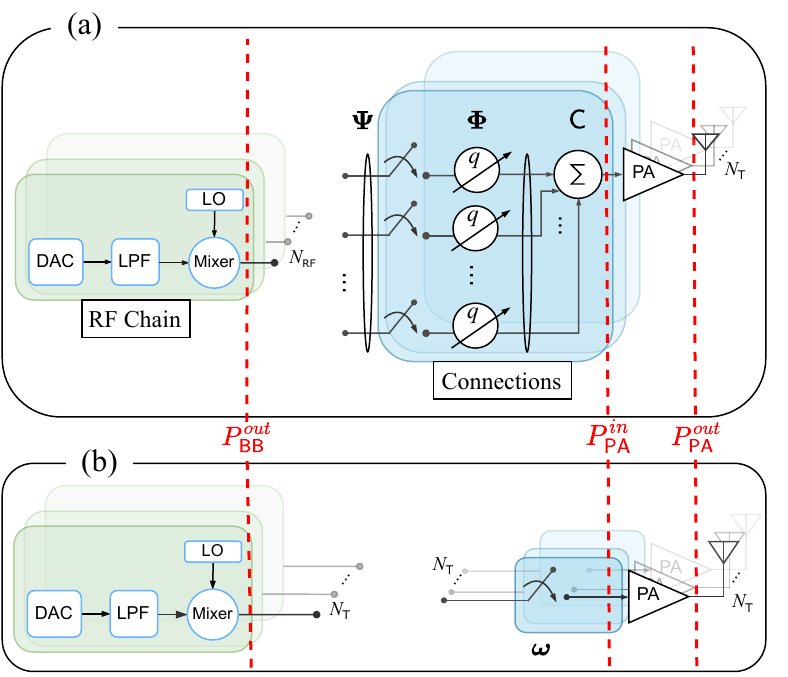}
    \caption{General beamforming structure. (a) Hybrid beamforming structure. (b) Fully digital precoder structure.}
    \label{fig:HBF_comp}
\end{figure}

\subsection{General Beamforming Structure}
We consider a regularity assumption where components of the same type have the same input/output interface, i.e. their inputs and outputs are connected to the same type and number of components. This assumption is generally true because it eases the conception of generic circuits. 


To better represent each \gls{HBF} structure, we suggest a general template form as shown in Figure~\ref{fig:HBF_comp} (a), where a given antenna is connected to a combiner having $c \in \{1,\ldots, N_{\sf{RF}}\}$ inputs. Each input of a combiner is connected to the output of a phase shifter. Then, each phase shifter is connected to an RF chain through a switch. The number of switches is $\psi \in \{1,\ldots,N_{\sf{RF}}\}$. 
As a result, the analog precoder can be fully characterized by specifying the tuple $(\psi,c)$. For instance, for the three conventional \gls{HBF} structures that we discussed previously, we have:
\begin{itemize}
    \item $(N_{\sf{RF}}, N_{\sf{RF}})$ for the FC-HBF structure. In the FC-HBF structure, all the switches are connected (i.e., $\psi = N_{\sf{RF}}$), while the outputs of all the PSs are combined before each antenna i.e., $c = N_{\sf{RF}}$.
    \item $(N_{\sf{RF}}, 1)$ for the DSA-HBF structure. In DSA-HBF, only one switch can be connected at each time interval, therefore $c = 1$, while there are possible connections for all the switches, thus $\psi = N_{\sf{RF}}$. It should be noted that such configuration for switches works like a multiplexer. Thus, in a practical system, the switches are replaced by a $\psi\times1$ multiplexer.
    \item $(1, 1)$ for the FSA-HBF structure. In FSA-HBF, each antenna is only connected to a fixed RF chain (i.e., $c = 1$), while the connection is fixed i.e., $\psi=1$.
\end{itemize}
The hardware complexity of different beamforming techniques is compared in Table~\ref{HW_Com}.

\begin{table}[t]
\centering
\caption{Energy Model Notations and Parameters}
\begin{tabular}{c|c|c}
\toprule
Component & Notation $\langle c\rangle$  &  Parameter $x$\\
\toprule
Digital analog convert & $\sf{D}$  & $\#$ bits $b_{D}$ \\
Low pass filter & $\sf{L}$  & - \\
Mixer & $\sf{M}$  & - \\
Local oscillator & $\sf{LO}$  & - \\
Switches (multiplexers) & $\bs{\Psi}$ & $\#$ inputs $\psi$ \\
Phase shifters & $\bs{\Phi}$  & -\\
Combiners & $\sf{C}$  & -\\
Power amplifier & $\sf{PA}$  & - \\
\bottomrule
\end{tabular}
\label{table:Energy_model}
\end{table}

\begin{table}[t!]
    \centering
    \caption{Hardware Complexity Comparison}
    \resizebox{\columnwidth}{!}{
  \begin{tabular}{c|ccccc}
    \toprule
      \multicolumn{1}{c}{Beamforming} &
      \multicolumn{5}{c}{Hardware components} \\
      {technique} & RF chains & Antennas & PS & Combiners & Switches \\
      \midrule
    FDP & $N_{\sf{T}}$ & $N_{\sf{T}}$ & - & - & -\\
    FC-HBF & $N_{\sf{RF}}$ & $N_{\sf{T}}$ & $N_{\sf{RF}}N_{\sf{T}}$ & $N_{\sf{T}}$ & -\\
    FSA-HBF & $N_{\sf{RF}}$ & $N_{\sf{T}}$ & $N_{\sf{T}}$ & - & -\\
    DSA-HBF & $N_{\sf{RF}}$ & $N_{\sf{T}}$ & $N_{\sf{T}}$ & - & $N_{\sf{T}}$ \\
    \bottomrule
  \end{tabular}}
  \label{HW_Com}
\end{table}

\subsection{Energy Consumption Analysis}
We now describe the energy consumption of each component, and we list the most recent state-of-the-art hardware solutions. We consider components that are suitable for operating in the frequency range of $20$-$40$ GHz. 

A component of the set $\{\sf{D}, \sf{L}, \sf{M}, \sf{LO}, \bs{\Psi}, \bs{\Phi},\sf{C},\sf{PA}\}$ is denoted by $o$ and the correspondence between a component and its notation is defined in Table~\ref{table:Energy_model}. We denote $\text{IL}_{o}$ as the insertion loss of passive component $o$ and when $o$ depends on some parameter $x$, we use $\text{IL}_{o}(x)$. The average power dissipated by the active component $o$ is denoted as $P_o$, or $P_{o}(x)$ if $o$ depends on the parameter $x$. See Table~\ref{table:Energy_model} for the list of components and their parameters. Note that the power dissipated by the wires is neglected and when $c = 1$ there is no need for a combiner (i.e., $\text{IL}_{\sf{C}}(1) = 0 \,\text{dB}$). Likewise, the switches can be replaced with wires when $\psi =1$ or $\psi= c$, that is $\text{IL}_{\bs{\Psi}}(1) = \text{IL}_{\bs{\Psi}}(c) = 0 \,\text{dB}$, since all possible connections are always established. 

In our energy model, we consider the possibility of turning off the RF chains or antennas to save power. The $n^{\text{th}}$ antenna or the $m^{\text{th}}$ RF chain is turned off when the $n^{\text{th}}$ row or the $m^{\text{th}}$ column of the matrix $\bs{\Omega}$ is zero, respectively. Therefore, we can define $N_{\sf{T}}(\bs{\Omega})=\{n: \sum_{m=1}^{N_{\sf{RF}}} [\mathbf{\Omega}]_{n,m}>0\}$ and $N_{\sf{RF}}(\bs{\Omega})=\{m:\sum_{n=1}^{N_{\sf{T}}} [\mathbf{\Omega}]_{n,m}>0\}$, as the set of activated antennas and RF chains, respectively.

\subsubsection{RF Front-End}
The RF front-end corresponds to the circuitry between the antenna and the \gls{DAC}. As shown in Figure~\ref{fig:HBF_comp}~(b), for the \gls{FDP}, this consists of low pass filters (LPFs), mixers, local oscillators (LOs), switches, and power amplifiers (PAs). On the other hand, in Figure~\ref{fig:HBF_comp}~(a), the \gls{HBF} requires a network of \gls{PS}s, splitters, and combiners in addition to the components described for the \gls{FDP}. Mixers, combiners, switches, and \glspl{PS} are assumed to be passive devices that introduce \gls{IL}. 

For the mixer, based on the recent solution in~\cite{9335489}, we consider $\text{IL}_{\sf{M}} = 6.4$~dB. The \gls{IL} of the \gls{PS} and the combiner plays a key role in designing energy-efficient \gls{HBF}, especially for the \gls{FC-HBF}, where all the RF chains are connected to all the antennas through \gls{PS}s and combiners. In Table~\ref{table:PS}, we list the \gls{IL}s of \gls{PS}s from some recent state-of-the-art references. Based on this table, we choose $\text{IL}_{\bs{\Phi}} = 3.7$~dB with $q=9.4$ bits resolution, and we assume $\text{IL}_{\sf{C}} = 1.8$~dB~\cite{comb1}. For \gls{DSA-HBF}, the switches dynamically change the connections between the RF chains and the antennas to improve the flexibility of the structure. Since these \glspl{IL} are in low power and they do not have a big impact on the final power consumption, we assume  $\text{IL}_{\bs{\Psi}}(\psi) = 1.1$~dB for the other values of $\psi$, by considering single pole single throw (SPST) switch~\cite{9443035}.

\begin{table}[!t]
\centering
\caption{Passive Phase Shifter IL Comparison}
\begin{tabular}{c|c|c|c|c}
\toprule
Reference & Year &   Frequency & bits & $\mathrm{IL}_{\bs{\Phi}}$\\
& & (GHz) & q & (dB)\\
\toprule
\cite{8781643} & 2019 & 24-28 & - & 5 \\
\cite{8617578} & 2018 & 28 & - & 4 \\
\cite{8432508} & 2018 & 22-36 & 3 & 5.6 \\
\cite{8541422} & 2018 & 28 & 9.4 & 3.7 \\
\bottomrule
\end{tabular}
\label{table:PS}
\end{table}

Now, denoting by $P_{\mathsf{BB}}^{out}$ the output power of each RF chain, the input power of the \gls{PA} before the $n^\text{th}$ antenna for all structures of the \gls{HBF} (in mW) can be written as
\begin{align} \label{general_P_HBF}
P_{\sf{PA}, \sf{HBF}}^{in,n} = \frac{P_{\mathsf{BB}}^{out}}{ \text{IL}_{\sf{C}}\text{IL}_{\bs{\Psi}}(\psi) \text{IL}_{\bs{\Phi}} \text{IL}_{\sf{M}}} \sum_{m \in N_{\sf{RF}}(\bs{\Omega})}\frac{[\bs{\Omega}_{\sf{HB}}]_{n,m}} {\sum_{n=1}^{N_{\sf{T}}}[\bs{\Omega}_{\sf{HB}}]_{n,m}},
\end{align} 
where $\text{IL}_{\bs{\Phi}}$ denotes the \gls{IL} of \glspl{PS} and \gls{IL} values are expressed in a linear scale. In the \gls{FC-HBF}, where all the RF chains are connected to the antennas ($\bs{\Omega}_{\sf{HB}}$ given in~\eqref{eq6}), we have $(\psi,c) = (N_{\sf{RF}}, N_{\sf{RF}})$ and $\text{IL}_{\bs{\Psi}}(\psi=c) = 1$. 
For \gls{DSA-HBF} with the structure of ($\psi$, c) = ($N_{\sf{RF}}, 1$) and the connection matrix $\bs{\Omega}_{\sf{HB}}$ in~\eqref{DSA_cons}, due to \gls{IL} of switches, we have $\text{IL}_{\bs{\Psi}}(\psi)~=~1.1$. In \gls{FSA-HBF} that has a structure $(\psi,c) = (1,1)$, there are neither combiners nor switches. As a result, $\text{IL}_{\bs{\Psi}}(1) = 1$, and $\text{IL}_{\sf{C}} = 1$.
Similarly for the \gls{FDP}, as shown in Figure~\ref{fig:HBF_comp}~(b), the input power of the \gls{PA} on the $n^{\text{th}}$ antenna can be obtained as
\begin{align}
P_{\sf{PA}, \sf{FDP}}^{in, n} =\frac{P_{\mathsf{BB}}^{out}}{ \text{IL}_{\sf{M}}}.
\end{align}
Finally, the direct current~(DC) power drawn by the $n^{\text{th}}$ active \gls{PA} $P_{\sf{PA}}$, can be written as
\beq \label{P_PA_BF}
P_{\sf{PA}, \sf{BF}}^{\textrm{DC}, n} =  \frac{P_{\sf{TX}}^{n} - P_{\sf{PA}, \sf{BF}}^{in, n}}{\alpha},
\eeq
where $\alpha$ is the power-added efficiency~(PAE) of the \gls{LPA}, $P_{\sf{TX}}^{n}$ is the transmit power of the $n^{\text{th}}$ antenna, 
and $\sf{BF}$ should be replaced with $\sf{HBF}$ or $\sf{FDP}$ according to the chosen transmitter type. Based on the recent solution for \gls{PA} listed in \cite{9223864,9338181,9369408}, we consider an average PAE  of $\alpha = 36$.

\subsubsection{Digital to Analog Converter}
\glspl{DAC} are among the components having the largest power consumption in wireless applications. The power consumed by a \gls{DAC} ($P_{\sf{D}}$) is a linear function of the sampling frequency ($f_{s}$) and the figure of merit~($\text{FoM}_{\sf{D}}$) of the converter, and grows exponentially with the number of bits of resolution ($b_{D}$) as 
$P_{\sf{D}} = \text{FoM}_{\sf{D}} \times f_s \times 2^{b_{D}}$~\cite{8883297}.
The sampling frequencies for ultra wide-band applications are in the range of $0.5$-$1$ GHz. It is shown in~\cite{8883297} that in terms of required signal-to-quantization noise ratio~(SQNR), \gls{FDP} required $2$ bits less than \gls{HBF}. Therefore, we assume $b_{D}=4$ for FDP and $b_{D}=6$ for HBF, respectively. Moreover, based on~\cite{9359109}, we consider $\text{FoM}_{\sf{D}} = 54.5$ fJ/conv.

\subsubsection{Low Pass Filter in TX}
The output of the DACs will require analog \gls{LPF} to reject spectral images and maintain out-of-band emission limits. For an $m^\prime$-th order active \gls{LPF} with cutoff frequency $f_{c}$, the $\text{FoM}_{\sf{L}}$ is the power consumed per pole per Hertz~\cite{6615940}. The power drawn by \gls{LPF} is given by
$P_{\sf{L}} = \text{FoM}_{\sf{L}} \times f_{c} \times m^\prime$.
Based on \cite{6615940}, we assume a first order \gls{LPF} with $f_{c} = 500 $MHz, and $\text{FoM}_{\sf{L}} = 1.4 \;\text{mW/GHz}$.
Furthermore, we define $P_{\sf{LO}}$ as the power consumed by the mixer from the \gls{LO} and we consider $P_{\sf{LO}} = 10$ dBm~\cite{6275452}.

\subsubsection{Total Energy Consumption}

Now, putting it all together, the total power consumed by a given beamforming structure can be written as follows:
\begin{align} \label{total_power1}
P_{\sf{BF}} = \big|N_{\sf{RF}}(\bs{\Omega})\big|( P_{\sf{L}}+ P_{\sf{LO}} + P_{\sf{D}}(b_{D})) + \sum_{n \in N_{\sf{T}}(\bs{\Omega})} P_{\sf{PA}, \sf{BF}}^{\textrm{DC}, n},  
\end{align}
where $P_{\sf{PA}, \sf{BF}}$ should be replaced with either $P_{\sf{PA}, \sf{HBF}}$ or $P_{\sf{PA}, \sf{FDP}}$ according to the transmitter type and $\bs{\Omega} \in \{\bs{\Omega}_{\sf{FD}},  \bs{\Omega}_{\sf{HB}}\}$.
In this paper, we focus on passive \gls{PS}, but we note that active \gls{PS} can be easily considered in the model by setting $\text{IL}_{\bs{\Phi}}$ to 1 and adding the power consumption of all active \glspl{PS} to \eqref{total_power1}. The energy consumption $E_{\sf{BF}}$ can then be obtained with $E_{\sf{BF}} = T_s \times P_{\sf{BF}}$, where $T_s$ is the duration of a symbol. When considering a fixed symbol duration, minimizing the power consumption is equivalent to minimizing the energy. Therefore, we evaluate the \gls{EE} as b/s/Hz/W. 
It is interesting to see that based on \eqref{general_P_HBF}, considering passive \glspl{PS} and combiners, the power consumed by different \gls{HBF} structures is similar since the \gls{IL} of the passive components is applied on the low power signals, before the \glspl{PA}. However, in terms of hardware complexity and cost, shown in Table~\ref{HW_Com}, the subarray \gls{HBF} is more efficient than \gls{FC-HBF}. 

In equations \eqref{eq:sumRate_HBF} and \eqref{total_power1}, we observe that both the \gls{SE} and the \gls{EE} are influenced by the matrix $\bs{\Omega}$. This matrix defines the connection between the RF chains and the antennas. Having more connections results in higher \gls{SE} as it increases beamforming flexibility. However, each connection corresponds to the use of an RF chain in \gls{FDP}, and in the case of \gls{HBF}, it involves a \gls{PS} and a combiner, leading to increased costs and energy consumption. This dependency makes the optimization problem in \eqref{SEE-max-prb} difficult to solve. Consequently, to address this issue, we propose a novel unsupervised learning solution in the following sections. This approach aims to jointly optimize both \gls{SE} and \gls{EE}.

\section{Energy-Efficient Beamforming Driven By Deep Unsupervised Learning} \label{sec:Optimal}

In this section, we describe the unsupervised learning solution to design the antenna selection and efficient HBF as well as \gls{FDP}. We start by describing the architecture of the proposed \gls{DNN} in Section~\ref{subsec:DNN}. Then, the proposed method is divided into two phases: the training phase is described in Section~\ref{subsec:training}, and the online phase is described in Section~\ref{subsec:online}. 

\begin{figure}[t!]
    \centering
    \includegraphics[width=\columnwidth]{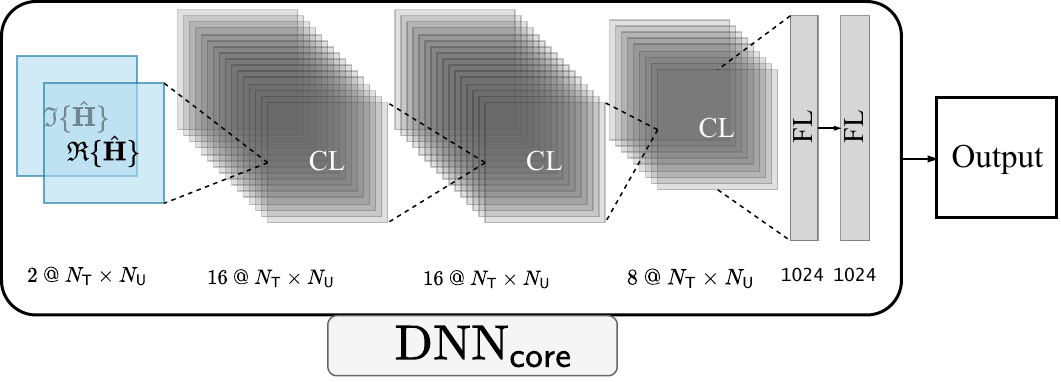}
    \caption{$\text{DNN}_{\sf{core}}$ architecture which is identical for both \gls{HBF} and \gls{FDP}.}
    \label{fig:DNN_core}
\end{figure}

\subsection{Deep Neural Network Architecture} \label{subsec:DNN}
The input and the hidden layers of the proposed DNN architecture are common for both the \gls{FDP} and the \gls{HBF} structures. However, the output layers are different for each BF structure. We start by describing the architecture of the input and the hidden layers denoted as $\text{DNN}_{\sf{core}}$ as shown in Figure~\ref{fig:DNN_core}. Then in the following subsections, we describe the architectures of the output layers of the \gls{HBF} and the \gls{FDP}.

The input of the \gls{DNN} is given by the imperfect channel matrix $\hat{\mb{H}}$ given in~\eqref{CSI_N}. To improve the representation learning, we normalize the input to $\bar{\mb{H}} = \hat{\mb{H}} / \|\hat{\mb{H}}\|_{F}^{2}$ such that $\|\bar{\mb{H}}\|_{F}^{2} = 1$. Then, we separate the real part $\Re\{\bar{\mb{H}}\}$ and the imaginary part $\Im\{\bar{\mb{H}}\}$ of $\bar{\mb{H}}$ into two channels that are fed to the first \gls{CL}.
$\text{DNN}_{\sf{core}}$ consists of $2$ \glspl{CL} $16 @ N_{\sf{T}} \times N_{\sf{U}}$ where $16$ is the number of channels and $N_{\sf{T}} \times N_{\sf{U}}$ is the dimension of each channel followed by $1$ CL $8 @ N_{\sf{T}} \times N_{\sf{U}}$. The kernel size is $3 \times 3$ for all \glspl{CL}. The \glspl{CL} are followed by $2$ \glspl{FL}, each with $1024$ neurons. The ``Leaky ReLU'' activation function and batch normalization are used after all layers except for the output layers.
This $\text{DNN}_{\sf{core}}$ is then combined with different output layers to form the HBF model, called 
efficient HBF network (E-HBF-Net), and the FDP model, called efficient FDP network (E-FDP-Net). 
The models are relatively small. For example, for $N_{\sf{T}}=64$, $N_{\sf{U}}=8$, the total number of parameters including the output layers in E-HBF-Net is $5.8$M (with $N_{\sf{RF}}=8$), and $4.9$M in E-FDP-Net.
A detailed complexity analysis is presented in Section~\ref{subsec:comp}.

\begin{figure}[t!]
    \centering
    \includegraphics[width=\columnwidth]{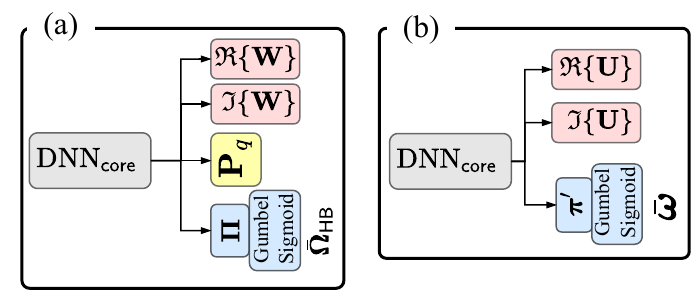}
    \caption{Proposed DNN architecture for (a) Hybrid beamforming, (b) Fully digital precoder.}
    \label{fig:DNN_arch}
\end{figure}

\subsubsection{Output Layers for HBF}
As shown in Figure~\ref{fig:DNN_arch}~(a), we divide the output of the last \gls{FL} into $4$ parallel layers.  The first and second parallel layers, both of size ${N_{\sf{RF}} \times N_{\sf{U}}}$, generate the real and imaginary part of the \gls{DP}. The output of the third parallel layer generates the AP, thus its dimension is $N_{\sf{RF}} \times N_{\sf{T}}$. The output of \gls{AP} can also be adapted to different \gls{PS} resolutions. 
It is shown in ~\cite{Flexible} that using the straight-through estimator~(STE) technique, we are able to have different numbers of quantization bits for the \glspl{PS}. In this paper we again consider the same approach for the output of the DNN dedicated for \gls{PS} quantization in \gls{AP}.
The fourth layer of size ${N_{\sf{RF}} \times N_{\sf{T}}}$ designs the matrix $\bs{\Omega}_{\sf{HB}}$. 

As we described before, $\bs{\Omega}_{\sf{HB}}$ must be a binary matrix. Typically, this binary constraint requires using the ``Sigmoid'' function during training and then, during the online phase, applying a rounding technique to transform the real values into binary values. However, we found that this approach does not lead to good results for unsupervised learning, because the \gls{SE} measured during training can be very different from the actual \gls{SE} measured during testing. To solve this problem, we propose to use a differentiable approximation, called ``Gumbel-Sigmoid'' during training inspired by the ``Gumbel-Softmax'' estimator~\cite{Gumbel}. The Gumbel-Softmax approximation is a technique that allows sampling from a categorical distribution during the forward pass of a neural network, by combining a re-parameterization trick and a smooth relaxation. The connection between the RF chains and the antennas can be represented using a categorical binary distribution. Hence, defining $\pi_{n,m}$ as the probability that antenna $n$ is connected to the RF chain $m$, then we can form an $N_{\sf{T}} \times N_{\sf{RF}}$ matrix $\bs{\Pi}$ that corresponds to the probability states between antenna $n$ and the RF chain $m$. The Gumbel-Softmax function, $G(\mb{\Pi})$, applied to each element of the matrix $\mb{\Pi}$ can then be defined as
\beq \label{Gumbel_Sig}
 \bar{\bs{\Omega}}_{\sf{HB}} = G(\bs{\Pi}) = \frac{\exp((\log(\bs{\Pi}) + \mb{g})/\tau)}{ \exp((\log(\bs{\Pi}) + \mb{g})/\tau) + \exp(\mb{g'}/\tau)},
\eeq
where $\bar{\bs{\Omega}}_{\sf{HB}}$ is the output of the DNN, and $\mb{g}$ and $\mb{g'}$ are independent samples with zero mean and unit variance, drawn from the Gumbel distribution. Note that the $\exp(\cdot)$ and $\log(\cdot)$ functions are applied element-wise when taking a matrix as input. The parameter $\tau$ is called the ``Gumbel temperature''. When $\tau \rightarrow 0$, $G(\bs{\Pi})$ tends to the categorical distribution, but when $\tau \rightarrow \infty$, it converges to the uniform distribution~\cite{Gumbel}. Therefore, there is a trade-off between small temperatures, where sample vectors are close to one-hot but the variance of the gradient is large, and large temperatures, where samples are more uniform but the variance of the gradient is small. We thus consider $\tau$ as a hyper-parameter to be optimized in our implementation.

\subsubsection{Output Layers for FDP}
The proposed architecture for \gls{FDP} is shown in Figure~\ref{fig:DNN_arch}~(b). We divide the output layer into 3 parallel layers. The first two layers are dedicated to the real and imaginary part of the \gls{FDP} with dimension $N_{\sf{T}} \times N_{\sf{U}}$. The third layer, similar to the one for \gls{HBF}, designs the antenna selection vector~($\bs{\omega}$) described in Section~\ref{subsub:FDP}. Here again, we use the Gumbel-Sigmoid described in \eqref{Gumbel_Sig} to obtain the binary variables from $\bs{\omega}$. Let $\pi'_{n}$ denotes the probability of activating the $n$-th antenna and $\bs{\pi}' = [\pi'_1, \ldots, \pi'_{N_{\sf{T}}}]$. Then, we have $\bar{\bs{\Omega}}_{\sf{FD}} = \text{diag}(\bar{\bs{\omega}})$, where $\bar{\bs{\omega}} = G(\bs{\pi}')$.

\begin{figure}[t!]
    \centering
    \includegraphics[width=\columnwidth]{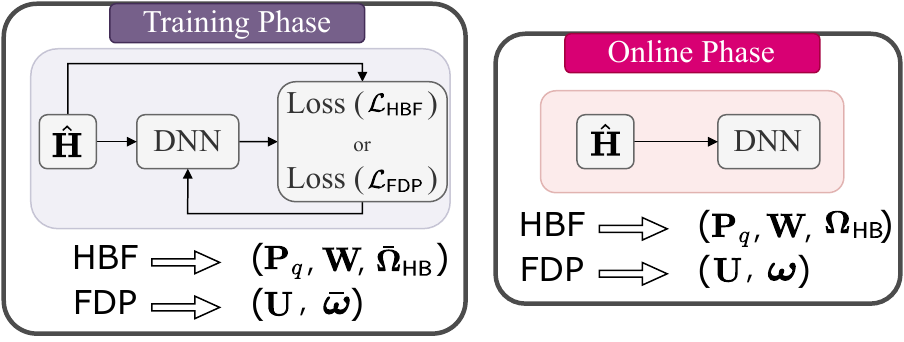}
    \caption{Training~(left) and online~(right) phases for efficient \gls{BF}. The outputs of the DNN depend on the BF structure~(HBF, FDP).}
    \label{fig:training_pro}
\end{figure}

\subsection{Training Phase: Unsupervised Learning}\label{subsec:training}

In the training phase, thanks to unsupervised learning, the data samples consist of only imperfect channel matrices without the need for labels. The imperfect channel~($\hat{\mb{H}}$) is modeled as in~\eqref{CSI_N} and it includes a coefficient $\beta$ that determines the magnitude of the estimation error and thus helps us study the impact of the estimation error of the channel on the DNN training. 

Although the approach to train the DNN is similar for E-HBF-Net and E-FDP-Net, there are differences in their hardware configurations. Therefore, we first present the common aspects shared by both DNN models and then proceed to explain the parts specific to each model.

The objective of the proposed solutions is to design the beamforming configuration to not only maximize the \gls{SE} but also to minimize the \gls{EC} while being adaptive to the number of active users, i.e., when the number of active users is small, it intelligently turns off part of the antennas since they will no longer be needed. Consequently, it will reduce the EC. To achieve this objective, we design the following unsupervised loss function to train the \gls{DNN}:
\beq \label{loss_setup}
\mathcal{L}_{\sf{BF}} = \gamma\mathcal{L}_{\sf{EC}}  + \mathcal{L}_{\sf{AAS}},
\eeq
where the first term is related to \gls{EC} and the second term is related to both the SE and the active number of users and is called the adaptive antenna selection (AAS) term. The hyper-parameter $\gamma$ is required to achieve proper training convergence and should be tuned in the training phase. Each term of the loss function is described in detail in the sequel.

EC  term ($\mathcal{L}_{\sf{EC}}$): This term is introduced to add a penalty to the total loss function to reduce \gls{EC}. It is given as:
\beq \label{loss_EE}
\mathcal{L}_{\sf{EC}} = \bar{P}_{\sf{BF}},
\eeq
where $\bar{P}_{\sf{BF}}$ is the total power consumption for either HBF~($\bar{P}_{\sf{HBF}}$) or FDP~($\bar{P}_{\sf{FDP}}$) given in~\eqref{total_power1} as discussed in Section~\ref{sec:Energ}, which depends on $\bar{\bs{\Omega}} \in \{\bar{\bs{\Omega}}_{\sf{FD}},  \bar{\bs{\Omega}}_{\sf{HB}}\}$. Thus, $\bar{\bs{\Omega}}_{\sf{BF}}$ affects both the \gls{SE} as well as the \gls{EC}.

AAS term ($\mathcal{L}_{\sf{AAS}}$): This term of the loss function $\mathcal{L}_{\sf{AAS}}$ is given by:
\beq \label{loss_AS}
\mathcal{L}_{\sf{AAS}} = \left(\frac{\bar{R}}{N_{\sf{U}}}- R_{\sf{d}}\right)^{2}, 
\eeq
where as discussed in Section~\ref{subsec:problem}, parameter $R_{\sf{d}}$ denotes the desired average \gls{SE} value for each user, and $\bar{R}$ is either $\bar{R}_{\sf{HBF}}(\bar{\mb{A}}, \mb{W})$ for HBF or $\bar{R}_{\sf{FDP}}(\mb{U} \times \bar{\bs{\Omega}}_{\sf{FD}})$ for FDP. Thanks to the AAS term, the SE is forced to approach $R_{\sf{d}}$ while parts of the antennas can be turned off to reduce the \gls{EC} (according to the EC term $\mathcal{L}_{\sf{EC}}$). As a result, the AAS term guarantees to consume minimum power to satisfy an average desired SE~($R_{\sf{d}}$).

\subsubsection{Efficient Hybrid Beamforming Network (E-HBF-Net)} 
To design an efficient \gls{HBF} structure, a programmable switch is employed for each connection ($N_{\sf{T}} \times N_{\sf{RF}}$) to find the best matrix ($\bs{\Omega}_{\sf{HB}}$) that maximizes the \gls{EE}.  As shown in the ``Training Phase'' of Figure~\ref{fig:training_pro}, the proposed DNN for HBF, E-HBF-Net, is designing jointly the \gls{DP}~($\mb{W} = \Re[\mb{W}] + i\Im[\mb{W}]$), the \glspl{PS}~($\bar{\mb{P}}_q$)\if represents the output of the \gls{DNN} for the phase shifters with a regression task\fi, and the connections matrix ($\bar{\bs{\Omega}}_{\sf{HB}}$) by employing the proposed ``Gumbel Sigmoid'' function as in \eqref{Gumbel_Sig}.

Obtaining $\bar{P}_{\sf{HBF}}$ requires computing~\eqref{total_power1} and thus we first need to know the power consumed by the \glspl{PA}. Thus, based on \eqref{P_PA_BF}, we would need the input and output power of the PAs. The output power of the PAs $1,\ldots,N_T$ is given by
\beq \label{pwr_hbf_norm2}
\bar{\mb{p}}_{\sf{TX}}=\sum_{u=1}^{N_{\sf{U}}}|\bar{\mb{A}}\mb{w}_u|^2,
\eeq
where $\bar{\mb{p}}_{\sf{TX}}=[\bar{P}_{\sf{TX}}^{1},\ldots,\bar{P}_{\sf{TX}}^{N_{\sf{T}}}]^{\sf{T}}$ and $\bar{\mb{A}}$ and $\mb{W}=[\mb{w}_1,\ldots,\mb{w}_{N_{\sf{U}}}]$ are the AP and DP outputs designed by the proposed DNN. Due to the total power constraint assumed at the BS in \eqref{cnt2}, we should normalize the power such $\|\bar{\mb{A}}\mb{W}\|_{F}^{2}=\sum_{n=1}^{N_{\sf{T}}}\sum_{u=1}^{N_{\sf{U}}}|\bar{\mb{A}}\mb{w}_u|^2 \leq P_{\sf{TX}}$. To respect the inequality of the power constraint, we introduce a new power threshold $\bar{P}_{\sf{TX}}$ that is a function of the connection matrix as follows:
\beq \label{pwr_AW}
\bar{P}_{\sf{TX}} = \sum_{\forall m,n} [\bar{\bs{\Omega}}_{\sf{HB}}]_{n,m}  P_{\sf{TX}} / (N_{\sf{RF}}N_{\sf{T}}).
\eeq
Therefore, the maximum transmitted power is limited to $P_{\sf{TX}}$ when all connections are established ($[\bar{\bs{\Omega}}_{\sf{HB}}]_{n,m} = 1 \; \forall n,m$), while reducing the number of connections reduces the transmit power. After power normalization, we can obtain the input power of the PAs according to~\eqref{general_P_HBF}. However, for the DNN loss function, we cannot have a sum over a dynamic set as defined in \eqref{general_P_HBF}. 
Therefore, we reformulate~\eqref{general_P_HBF} as
\begin{align} \label{pwr_in_PAs_training}
\bar{\mb{p}}_{\sf{PA}, \sf{HBF}}^{in} & = \frac{P_{\mathsf{BB}}^{out}}{ \text{IL}_{\sf{C}}\text{IL}_{\bs{\Psi}}(\psi) \text{IL}_{\bs{\Phi}} \text{IL}_{\sf{M}}}\bar{\bs{\Omega}}_{\sf{HB}}\text{diag}(\bar{\bs{\Omega}}_{\sf{HB}}^{\rm{T}}\bs{1}_{N_{\sf{T}}})^{\dagger}\bs{1}_{N_{\sf{RF}}},
\end{align}
where $\bar{\mb{p}}_{\sf{PA}, \sf{HBF}}^{in} = [\bar{P}_{\sf{PA}, \sf{HBF}}^{in,1}, \ldots, \bar{P}_{\sf{PA}, \sf{HBF}}^{in, N_{\sf{T}}}]$ is the vector of input power of the \glspl{AP}, $\bar{\bs{\Omega}}_{\sf{HB}}$ is the output of Gumbel-Sigmoid function for HBF, and $\bs{1}_{N}$ denotes the all-one column vector of size $N$. According to~\eqref{P_PA_BF} and~\eqref{pwr_hbf_norm2}-\eqref{pwr_in_PAs_training}, we can obtain $\bar{P}_{\sf{PA}, \sf{HBF}}^{\textrm{DC}, n}$.

To compute the power consumption of all activated RF chains as in~\eqref{total_power1}, we need to determine the number of activated RF chains (i.e., $N_{\sf{RF}}(\bar{\bs{\Omega}}_{\sf{HB}})$). However, finding $N_{\sf{RF}}(\bar{\bs{\Omega}}_{\sf{HB}})$ again requires a summation over a dynamic set and it is not appropriate for the loss function. As a consequence, we use an alternative linear algebra formulation. First, we compute the expectation over all antennas of each RF chain as $\bar{\bs{\Omega}}_{\sf{HB}}^{\rm{T}}\bs{1}_{N_{\sf{T}}}/N_{\sf{T}}$. Then, we find the expected number of activated RF chains as follows:
\beq \label{car_rf}
\bar{N}_{\sf{RF}}(\bar{\bs{\Omega}}_{\sf{HB}}) = \bs{1}_{N_{\sf{RF}}}^{\rm{T}}\bar{\bs{\Omega}}_{\sf{HB}}^{\rm{T}}\bs{1}_{N_{\sf{T}}}/N_{\sf{T}}.
\eeq
Finally, the total power consumption is given by
\begin{align} \label{total_power}
\bar{P}_{\sf{HBF}} = \bar{N}_{\sf{RF}}(\bar{\bs{\Omega}}_{\sf{HB}})( P_{\sf{L}}+ P_{\sf{LO}} + P_{\sf{D}}(b_{D})) + \sum_{n = 1}^{N_{\sf{T}}} \bar{P}_{\sf{PA}, \sf{HBF}}^{\textrm{DC}, n},  
\end{align}
where $\bar{P}_{\sf{HBF}}$ is the power consumption terms that has been employed in \eqref{loss_EE}.

\algnewcommand{\algorithmicgoto}{\textbf{go to}}%
\algnewcommand{\Goto}[1]{\algorithmicgoto~\ref{#1}}%
{\SetAlgoNoLine%
\begin{algorithm}[t] \label{alg_HBF_EE}
\hspace*{\SpaceReservedForComments}{}%
\begin{minipage}{\dimexpr\linewidth-\SpaceReservedForComments\relax}
  \caption{Efficient HBF (E-HBF-Net)}
  \begin{algorithmic}[1]
    \State \textbf{Input:} $\bar{\mb{H}}$
    \State \textbf{Output:} $\Re[\mb{W}]$, $\Im[\mb{W}]$, $\mb{P}_{q}$, and $\bar{\bs{\Omega}}_{\sf{HB}}$
    \State \textbf{Hyper-Parameters: } $\gamma$, $R_{\sf{d}}$
    \State \For{$i$ \textbf{in} epochs}
    {\State \quad \textbf{FeedForward} E-HBF-Net.train()
    \State \quad $\mb{W} = \Re[\mb{W}] + i\Im[\mb{W}]$
    \State \quad $\mb{A} = \mb{P}_{q} \otimes \bar{\bs{\Omega}}_{\sf{HB}},$
    \State \quad $\bar{P}_{\sf{TX}}^{n}$, $\bar{P}_{\sf{PA}, \sf{HBF}}^{in, n}$, and $\bar{N}_{\sf{RF}}(\bar{\bs{\Omega}}_{\sf{HB}})$ in \eqref{pwr_hbf_norm2} in \eqref{pwr_in_PAs_training}, and  \eqref{car_rf}
    
    \State \quad \textbf{compute} $P_{\sf{HBF}}$ based on \eqref{total_power}
    \State \quad \textbf{compute} $R_{\sf{HBF}}(\bar{\mb{A}},\mb{W})$ based on \eqref{eq:sumRate_HBF}
    \State \quad \textbf{Loss:} $\mathcal{L}_{\sf{BF}} =\gamma\mathcal{L}_{\sf{EC}}+ \mathcal{L}_{\sf{AAS}}$ 
    \State \quad \textbf{compute}  gradient over layers
    \State \quad \textbf{update} weights and biases with \textbf{AdamW} optimizer
    }
    \State \textbf{Input:} $\hat{\mb{H}}$
    \State \textbf{Output:} $\mb{W}$, $\mb{P}_{q}$, and $\bs{\Omega}_{\sf{HB}}$
    \State \textbf{FeedForward} E-HBF-Net.eval()
    \State $\bs{\Omega}_{\sf{HB}} = \lfloor \bar{\bs{\Omega}}_{\sf{HB}} \rceil$
    \State $\mb{W} = \Re[\mb{W}] + i\Im[\mb{W}]$
    \State $\mb{A} = \mb{P}_{q} \otimes \bs{\Omega}_{\sf{HB}},$
\end{algorithmic}%
\AddNote[black]{1}{13}{Training Phase}
\AddNote[blue]{14}{19}{Online Phase}
\end{minipage}%
\end{algorithm}} 

{\SetAlgoNoLine%
\begin{algorithm}[t] \label{alg_FDP_EE}
\hspace*{\SpaceReservedForComments}{}%
\begin{minipage}{\dimexpr\linewidth-\SpaceReservedForComments\relax}
  \caption{Efficient FDP (E-FDP-Net)}
  \begin{algorithmic}[1]
    \State \textbf{Input:} $\hat{\mb{H}}$
    \State \textbf{Output:} $\Re[\mb{U}]$, $\Im[\mb{U}]$, and $\bar{\bs{\Omega}}_{\sf{FD}}$
    \State \textbf{Hyper-Parameters: } $\gamma$, $R_{\sf{d}}$
    \State \For{$i$ \textbf{in} epochs}
    {\State \quad \textbf{FeedForward} E-FDP-Net.train()
    \State \quad  $\bar{\bs{\Omega}}_{\sf{FD}} = \text{diag} (\bar{\bs{\omega}})$ 
    \State \quad \textbf{compute} $\bar{P}_{\sf{TX}}^{n}$ as \eqref{FDP_car_nt}
    \State \quad $\bar{\mb{U}} = \Re[\mb{U}] + i\Im[\mb{U}]$
    \State \quad \textbf{compute} $P_{\sf{FDP}}$ based on \eqref{total_power}
    \State \quad \textbf{compute} $R_{\sf{FDP}}(\bs{\bar{\Omega}}_{\sf{FD}} \times \mb{U})$ based on \eqref{eq:FDP_SINR}
    \State \quad \textbf{Loss:} $\mathcal{L}_{\sf{BF}} = 
    \gamma\mathcal{L}_{\sf{EC}}+\mathcal{L}_{\sf{AAS}}$ 
    \State \quad \textbf{compute}  gradient over layers
    \State \quad \textbf{update} weights and biases with \textbf{AdamW} optimizer
    }
    \State \textbf{Input:} $\hat{\mb{H}}$
    \State \textbf{Output:} $\mb{U}$, and $\bs{\Omega}_{\sf{FD}}$
    \State \textbf{FeedForward} E-FDP-Net.eval()
    \State $\bs{\Omega}_{\sf{FD}} = \text{diag} (\lfloor \bar{\bs{\omega}} \rceil)$
    \State $\mb{U} = \Re[\mb{U}] + i\Im[\mb{U}]$
\end{algorithmic}%
\AddNote[black]{1}{13}{Training Phase}
\AddNote[blue]{14}{18}{Online Phase}
\end{minipage}%
\end{algorithm}}

\subsubsection{Fully Digital Precoder (E-FDP-Net)}
E-FDP-Net provides the precoder $\mb{U} = \Re[\mb{U}] + i\Im[\mb{U}]$ and the vector $\bar{\bs{\omega}}$ for antenna selection, where $\bar{\bs{\Omega}}_{\sf{FD}} = \text{diag} (\bar{\bs{\omega}})$. To evaluate the first term of the loss function detailed in~\eqref{loss_setup}, the total power consumption of FDP~($\bar{P}_{\sf{FDP}}$) is required. 
Computing $\bar{P}_{\sf{FDP}}$ for E-FDP-Net is simpler than HBF because in FDP each antenna is connected to one RF chain. Consequently, the input power of each PA is simply $P_{\sf{BB}}^{out}$. Similar to HBF, to respect the power constraint for FDP, $\|\bar{\mb{U}}\|_{F}^{2} \leq P_{\sf{TX}}$, the output power should be a function of $\bar{\bs{\omega}} = [\bar{\omega}_1, ..., \bar{\omega}_{N_{\sf{T}}}]$. As a consequence, we denote the output power of the $n^{\text{th}}$ antenna as
\beq \label{FDP_car_nt}
\bar{P}_{\sf{TX}}^{n} = \sum_{u=1}^{N_{\sf{U}}}|[\mb{U}]_{n,u}|^2\bar{\omega}_n \,.
\eeq
Therefore, the power consumed by the PAs is given by~\eqref{P_PA_BF}.
Finally, the power consumed by the active RF chains is also easy to compute because the number of active RF chains is given by $\bar{N}_{\sf{RF}}(\bar{\bs{\Omega}}_{\sf{FD}}) = \sum_{n=1}^{N_{\sf{T}}} \bar{\bs{\omega}}$.



\subsection{Online Phase: Transmitting Data}\label{subsec:online}

Once the DNN has been trained, the online phase can start as shown in Figure~\ref{fig:training_pro}~(right). In the online phase, the \gls{DNN} input is only given by the imperfect channel matrices~$\hat{\mb{H}}$. In the online phase, like the training phase, the outputs of the DNN: the AP ($\mb{P}_q$) and the DP ($\mb{W}$) in \gls{HBF} and ($\mb{U}$) in FDP can be employed as is without any further processing, which is not the case for the connection matrix $\bar{\bs{\Omega}}$. Since the connection matrix ($\bar{\bs{\Omega}}_{\sf{HB}}$ in \gls{HBF} or $\bar{\bs{\omega}}$ in \gls{FDP}) should be binary, once it is output by the DNN in the online phase, it requires binary quantization. To do so, we can use the element-wise round function~($\lfloor\cdot\rceil$) on each element of the connection matrix as follows: $\bs{\Omega}_{\sf{HB}} = \lfloor \bar{\bs{\Omega}}_{\sf{HB}} \rceil$ for \gls{HBF} and $\bs{\omega} = \lfloor \bar{\bs{\omega}} \rceil$, and $\bs{\Omega}_{\sf{FD}} = \text{diag}(\bs{\omega})$ for \gls{FDP}. The output power of the $n^{\text{th}}$ antenna for E-HBF-Net is given by the $n^{\text{th}}$ element of the power vector defined in~\eqref{pwr_hbf_norm2} \if$P_{\sf{TX}}^{n} = [\sum_{u=1}^{N_{\sf{U}}}|\bar{\mb{A}}\mb{w}_u|^2]_n,$\fi
while for E-FDP-Net it is given by~\eqref{FDP_car_nt}. \if$P_{\sf{TX}}^{n} = \sum_{u=1}^{N_{\sf{U}}}|[\mb{U}]_{n,u}|^2\omega_n.$\fi
The two proposed DNN solutions, E-HBF-Net and E-FDP-Net, are described in Algorithm~\ref{alg_HBF_EE}, Algorithm~\ref{alg_FDP_EE}, respectively.

\section{Performance Evaluation} \label{sec:sim}

\begin{figure}[t!]
    \centering
    \includegraphics[width=0.9\columnwidth]{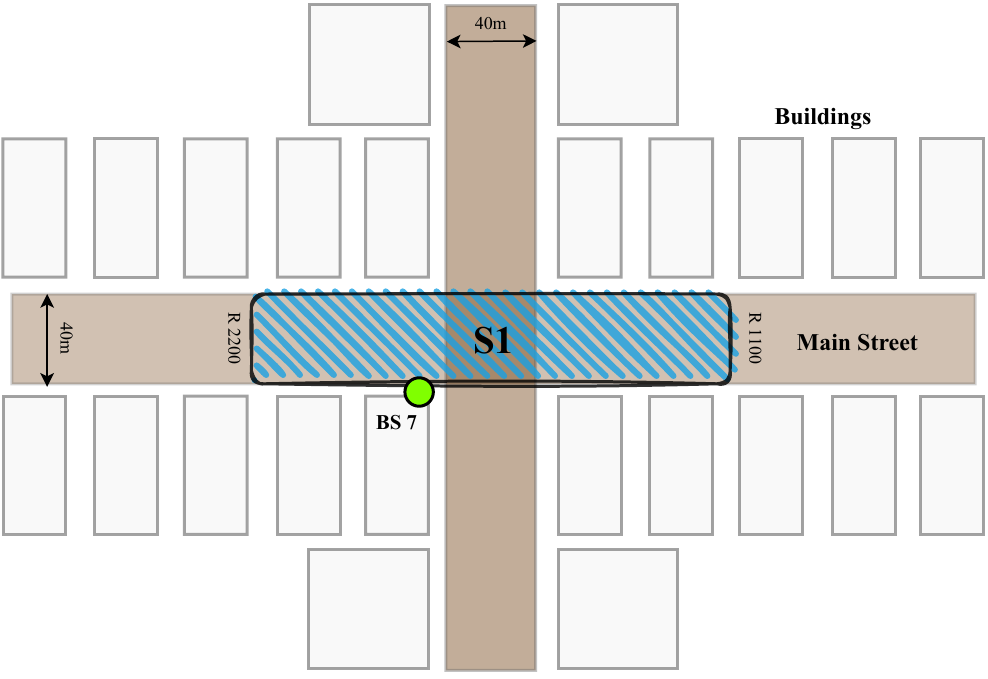}
    \caption{Illustration of the selected covered area from deepMIMO channel model~\cite{deepmimo}.}
    \label{fig:deepmimo}
\end{figure}

In this section, the performance of the proposed DNN, implemented using the \textsc{PyTorch} deep learning library, is numerically evaluated. The scenario ``O1-$28$~GHz'' of the deepMIMO channel model~\cite{deepmimo} is employed to generate the unlabeled dataset (the channel coefficients $\mb{h}_{u}$ for user $u$) for the training and testing. In the deepMIMO dataset~\cite{deepmimo}, realistic channel information is generated by applying ray-tracing methods to a three-dimensional model of an urban environment to capture the geometry-based characteristics, such as the correlation between the channels at different locations, and the dependence on the materials of the various environmental elements, among others. The parameters to generate the deepMIMO dataset are shown in Table~\ref{tab:deepMIMO_params}, where the channel model parameters $\small\texttt{active\_user\_first}$ and $\small\texttt{active\_user\_last}$ are set to 1100 and 2200 respectively. The \gls{BS} is equipped with $N_{\sf{T}} = 64$ antennas and $N_{\sf{RF}} = 8$ RF chains with \glspl{PS} serving $N_{\sf{U}}=4$ users randomly located in a dedicated area~(\textbf{S1} in Figure~\ref{fig:deepmimo}). Scenario ``O1'' consists of several users' locations being randomly placed in two streets surrounded by buildings. These two streets are orthogonal and intersect in the middle of the considered area. The size of the DNN dataset is set to $2 \times 10^{6}$ samples, with $85$\% of the samples used for the training set and the remaining used to evaluate the performance. We used ``AdamW'' as the DNN training optimizer. The hyper-parameters used in our DNN model are listed in Table~\ref{tbl:HP-DNN}.  In addition, hyper-parameter $\tau$ known as the Gumbel-Sigmoid temperature is set to $0.1$ and $0.5$ for E-HBF-Net and E-FDP-Net, respectively, while the best value for hyper-parameter $\gamma$, described in \eqref{loss_setup}, depends on $R_{\sf{d}}$, and ranges from $\gamma=0.1$ for $R_{\sf{d}}=1$, to $\gamma=0.005$ for $R_{\sf{d}}=8$. 
The training procedure required $200$ epochs.

\begin{table}[t]
\centering
\caption{\vspace{1pt}Parameter selection for the deepMIMO channel model}
\vspace{-4pt}
\resizebox{0.8\columnwidth}{!}{
\begin{tabular}{p{1.5cm}c|p{2cm}c}
\toprule
\multicolumn{2}{c|}{System}  & \multicolumn{2}{c}{Antennas} \\
Parameter           &  Value & Parameter   &  Value \\
\midrule
\texttt{scenario}  & ``O1-28 GHz''     & \texttt{num\_ant\_x} & $1$      \\
\texttt{bandwidth} & $0.5$ GHz    & \texttt{num\_ant\_y} & $8$    \\
\texttt{num\_OFDM} & $512$   & \texttt{num\_ant\_z} & $8$      \\
\texttt{num\_paths} & $2$      & \texttt{ant\_spacing} & $0.5$    \\
\bottomrule
\end{tabular}}
\label{tab:deepMIMO_params}
\end{table}

\begin{table}[t]
    \centering
    \caption{Proposed DNN hyper-parameters}
    \resizebox{0.7\columnwidth}{!}{
    \begin{tabular}{lc}
        \toprule
        \multicolumn{1}{c}{Parameter} & \multicolumn{1}{c}{Set Value} \\
        \cmidrule(lr){1-1} \cmidrule(lr){2-2}         
        Mini-batch size& 1000 \\
        Initial learning rate &  0.005 \\
        ReduceLROnPlateau (factor) & 0.4 \\
        ReduceLROnPlateau (patience) & 10 \\
        Weight decay & $10^{-5}$ \\ 
        Dropout keep probability & .95 \\
        Kernel size & 3 \\
        Zero padding & 1 \\
        ``$\epsilon$'' in BatchNorm (1D \& 2D) & $10^{-5}$\\
        \bottomrule
    \end{tabular}}
    \label{tbl:HP-DNN}
\end{table}

\begin{table}[t]
\centering
\caption{Simulation parameters for the energy model}
\resizebox{0.7\columnwidth}{!}{
 \begin{tabular}{cc|cc} 
 \toprule
 \multicolumn{1}{c}{Parameter} & \multicolumn{1}{c}{Value} & \multicolumn{1}{c}{Parameter} & \multicolumn{1}{c}{Value} \\
 \cmidrule(lr){1-2} \cmidrule(lr){3-4}
 $\text{FoM}_{\sf{L}}$ & 1.4 mW/GHz & $\text{IL}_{\sf{C}}$ & $1.8$ dB\\ [.2ex]
 $\text{FoM}_{\sf{D}}$ & $54.5$ fJ/conv & $\text{IL}_{\sf{M}}$ & $5.5$ dB\\ [.2ex]
 $f_s$ & $0.5$ GHz& $\text{IL}_{\bs{\Phi}}$ & $3.7$ dB\\ [.2ex]
 $f_{c}$ & $500$ MHz & $\text{IL}_{\bs{\Psi}}$ & $1$ dB\\ [.2ex]
 $P_{\sf{TX}}$ & $40$ dBm & $P_{\sf{LO}}$& $10$ mW\\ [.2ex]
 $\alpha$ & $36$\%  & $P_{\sf{L}}$ & $0.7$ mW\\[.2ex]
 $P_{\sf{BB}}^{out}$ & $-5.6$ dBm & $b_D$ & 6 bits\\[.5ex]
 \hline
 \end{tabular}}
 \label{tbl:EM_Par}
 \end{table}

\subsection{Spectral Efficiency and Power Consumption Analysis}

\begin{figure}[t!]
    \centering
    \includegraphics[width=\columnwidth]{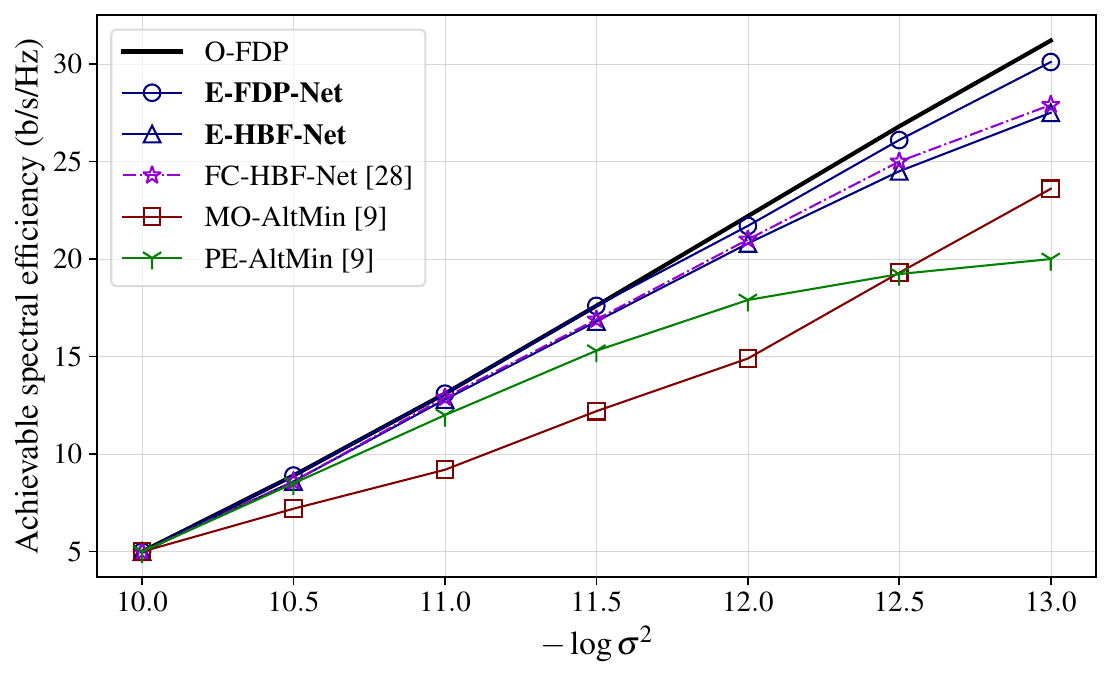}
    \caption{Maximum spectral efficiency of E-FDP-Net and E-HBF-Net, compared to other conventional approaches. System parameters are set to: $N_{\sf{U}} = 4$, $N_{\sf{RF}} = 8$, $N_{\sf{T}} = 64$.}
    \label{res:SR}
\end{figure}

We first verify the maximum SE that can be achieved by the proposed DNNs, when they are trained without considering their power consumption, and compare them with the baseline solutions presented in Section~\ref{Sec:Baseline}.
This maximal SE is shown in Figure~\ref{res:SR} when varying the noise power. Taking into account channel attenuation, the average \gls{SNR} ranges from $-7.8$ dB to $22.2$ dB. 
To obtain the maximum SE, we set $\gamma = 0$ so that the loss function for E-HBF-Net and E-FDP-Net in~\eqref{loss_EE} depends only on $\mathcal{L}_{\sf{AS}}$ and we set $R_{\sf{d}}=15$ to have no constraint on SE. On the one hand, the proposed E-FDP-Net gives a close-to-optimal performance. On the other hand, E-HBF-Net, outperforms other conventional solutions and is very close to E-FDP-Net performance. In the low-noise regime, the \gls{SE} of all solutions continues to increase. However, both E-HBF-Net and E-FDP-Net outperform other conventional non-DL solutions in high SNR regimes.

\begin{figure}[t!]
    \centering
    \includegraphics[width=\columnwidth]{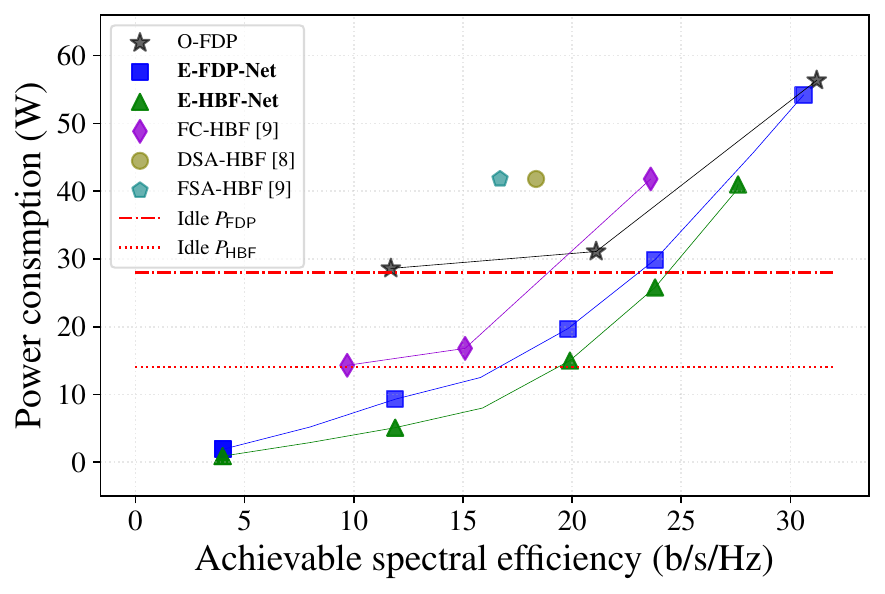}
    \caption{Power required to achieve a given SE for the various transmitter configurations. Idle $P_{\sf{BF}}$ is the power consumed by the BF structure when $P_{\sf{TX}}=0$. The parameters are set to: $N_{\sf{U}} = 4$, $N_{\sf{T}} = 64$, $N_{\sf{RF}} = 8$, and $\sigma^2 = -130$ dBm.}
    \label{res:scat}
\end{figure}
In Figure~\ref{res:scat}, we compare the power consumption of different BF hardware configurations at a given SE. It is shown that by adjusting $R_{\sf{d}}$ for E-FDP-Net and E-HBF-Net, different SE and power consumption trade-offs can be obtained, where for each proposed technique we set $R_{\sf{d}}$ in $\{1, 3, 5, 6, 8\}$. To cover a range of SE values, we also adjust the transmitted power for the conventional methods by setting $P_{\sf{TX}}$ in $\{0.1, 1, 10\}$W.
We see that the optimal FDP and the proposed E-FDP-Net with $R_{\sf{d}} = 8$ achieve the best SE. However, they also consume the most power because they require to activate all $N_{\sf{T}}$ RF chains. 
In this figure, we see that when the desired SE parameter $R_d$ is reduced, both E-FDP-Net and E-HBF-Net are able to reduce their power consumption. For example, when $R_{\sf{d}}$ is decreased from $8$ to $5$ bits/s/Hz/user, the consumed power for both E-FDP-Net and E-HBF-Net is reduced significantly ($64\%$ less for E-FDP-Net and $68\%$ less for E-HDF-Net).
By decreasing $R_{\sf{d}}$ further, both the power consumption and the SE continue to decrease. 
Furthermore, we see that E-FDP-Net and E-HBF-Net achieve must better energy efficiency than the baseline approaches.
For example, when $R_{\sf{d}} = 6$, it can be seen that E-HBF-Net achieves similar SE compared to FC-HBF solved with MO-AltMin, 
but with almost $1.7$ times less consumed power.
Further, the baseline solutions exhibit a power floor, shown by red lines in the figure, that corresponds to the power consumed by RF chains. When the transmit power of O-FDP and FC-HBF is decreased to $P_{\sf{TX}} = 1$W and $P_{\sf{TX}} = 0.1$W, the SE is degraded due to the lower transmit power. However, there is constant power consumption for each beamforming technique due to the operation of RF chains. 
On the contrary, E-FDP-Net and E-HBF-Net have the ability to reduce their power consumption below these floors by adaptively turning off their RF chains.

\begin{figure}[t!]
    \centering
    \includegraphics[width=\columnwidth]{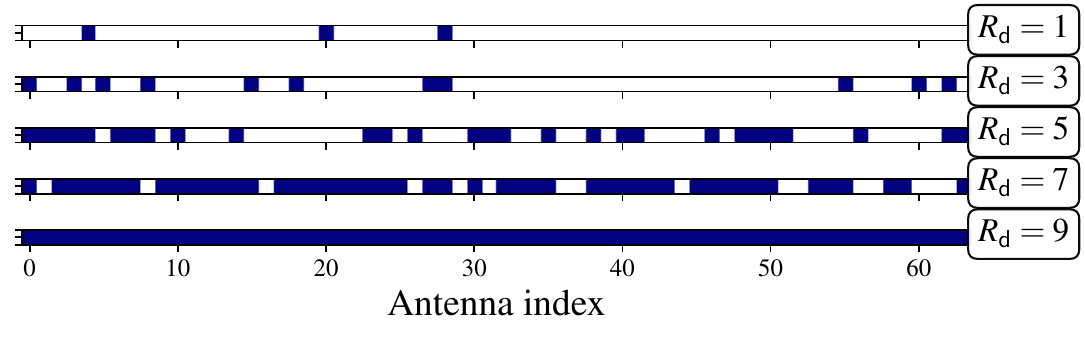}
    \caption{The connection matrix $\bar{\bs{\Omega}}_{\sf{FD}} = \text{diag}(\bar{\bs{\omega}})$ of E-FDP-Net for one input sample, and for different values of hyper-parameter $R_{\sf{d}}$, where a blue square represents the value $1$ and a white one represents the value $0$. System parameters are set to: $N_{\sf{U}} = 4$, $N_{\sf{T}} = 64$, and $\sigma^2 = -130$ dBm.}
    \label{res:gamma_prim}
\end{figure}

To illustrate how many antennas are activated by E-FDP-Net, we plot in Figure~\ref{res:gamma_prim} the connection matrix $\bar{\bs{\Omega}}_{\sf{FD}}$ for one sample of the test set, for different values of $R_{\sf{d}}$, where a blue square represents the value $1$ and a white square represents the value $0$. It can be seen that large values of $R_{\sf{d}}$ lead to more active antennas (and thus more active RF chains), and thus to a higher power consumption. 
\begin{figure}[t!]
    \centering
    \includegraphics[width=\columnwidth]{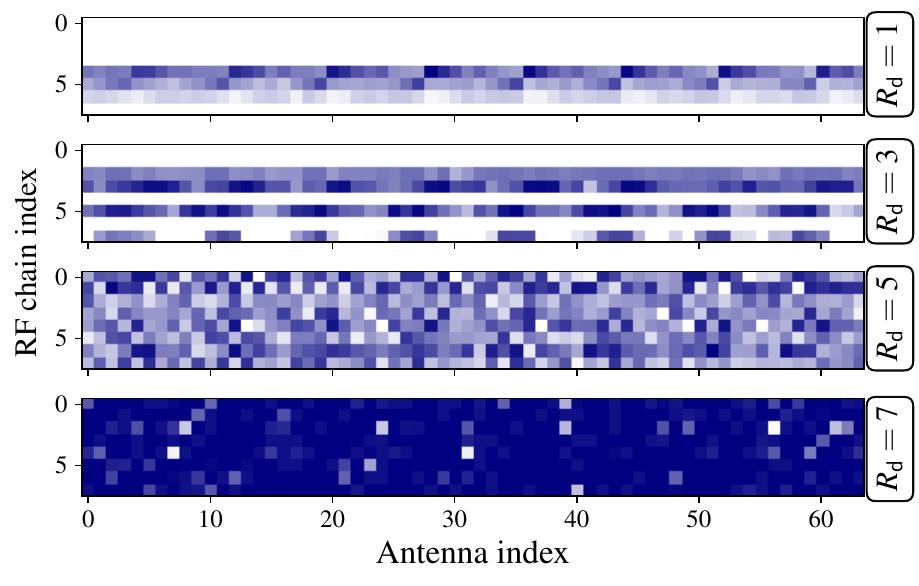}
    \caption{The average value of the connection matrix $\bar{\bs{\Omega}}_{\sf{HB}}$ of E-HBF-Net given for different values of hyper-parameter $R_{\sf{d}}$, where the shade of each square represents the range of values from $0$ (light) to $1$ (dark). System parameters are set to: $N_{\sf{U}} = 4$, $N_{\sf{T}} = 64$, $N_{\sf{RF}} = 8$, and $\sigma^2 = -130$ dBm.}
    \label{res:gamma}
\end{figure}
In Figure~\ref{res:gamma}, we show the average value of  $\bar{\bs{\Omega}}_{\sf{HB}}$ over the inputs, for different values of $R_{\sf{d}}$. When decreasing $R_{\sf{d}}$, the number of active antennas (non-zero columns) remains constant, while the number of active RF chains (non-zero rows) is reduced. 
This is because the power consumption of an antenna depends on its transmit power, which can be adjusted, whereas RF chains consume a fixed amount of power and must be turned off to save power.
It is interesting to see that with a lower value of $R_{\sf{d}}$, the E-HBF-Net designs the connection matrix such that a small number of RF chains are activated that are connected to several antennas, which helps to increase the spatial multiplexing gain and degrees of freedom. Finally, we see that as $R_{\sf{d}}$ increases, more antennas and more RF chains are activated, and thus more power is used.

\begin{figure}[t!]
    \centering
    \includegraphics[width=\columnwidth]{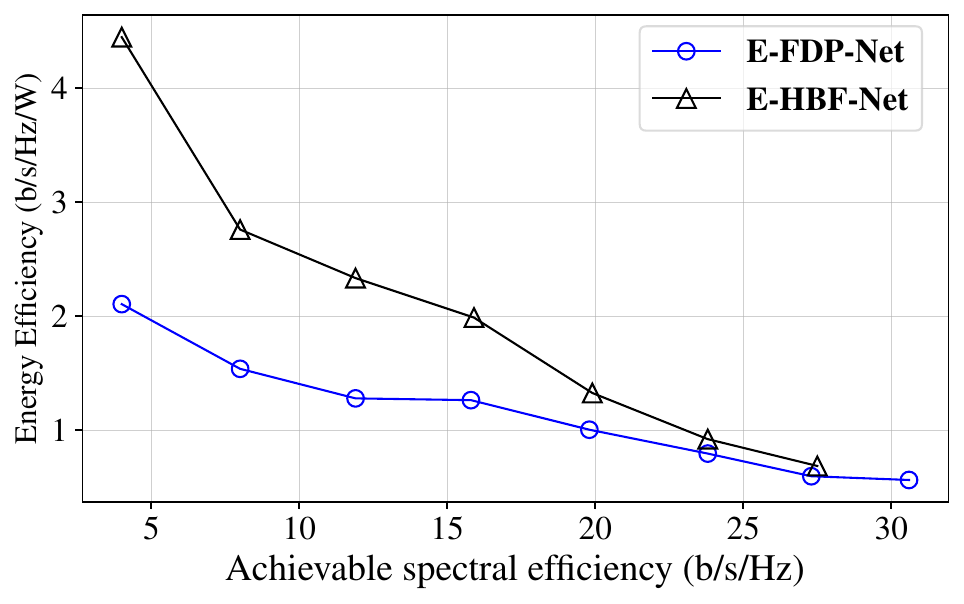}
    \caption{EE versus SE for the proposed E-HBF-Net and E-FDF-Net. The parameters are set to: $N_{\sf{U}} = 4$, $N_{\sf{T}} = 64$, $N_{\sf{RF}} = 8$, and $\sigma^2 = -130$ dBm.}
    \label{res:EE_SR}
\end{figure}

Figure~\ref{res:EE_SR} presents the EE versus SE comparison for the proposed E-FDP-Net and E-HBF-Net, with varying adjustments to $R_{\sf{d}}$. Notably, as SE decreases, E-HBF-Net demonstrates superior EE performance compared to E-FDP-Net. This outcome is attributed to the behavior of E-HBF-Net at lower SE values, where it intelligently deactivates RF chains while keeping multiple antennas active. Conversely, in E-FDP-Net, turning off an RF chain also turns off the associated antenna. Consequently, E-HBF-Net excels in conserving energy while simultaneously offering enhanced SE due to its higher flexibility. Furthermore, as SE increases, E-HBF-Net maintains its efficiency advantage over E-FDP-Net, although the performance gap between the two approaches diminishes.


\begin{figure}[t!]
    \centering
    \includegraphics[width=\columnwidth]{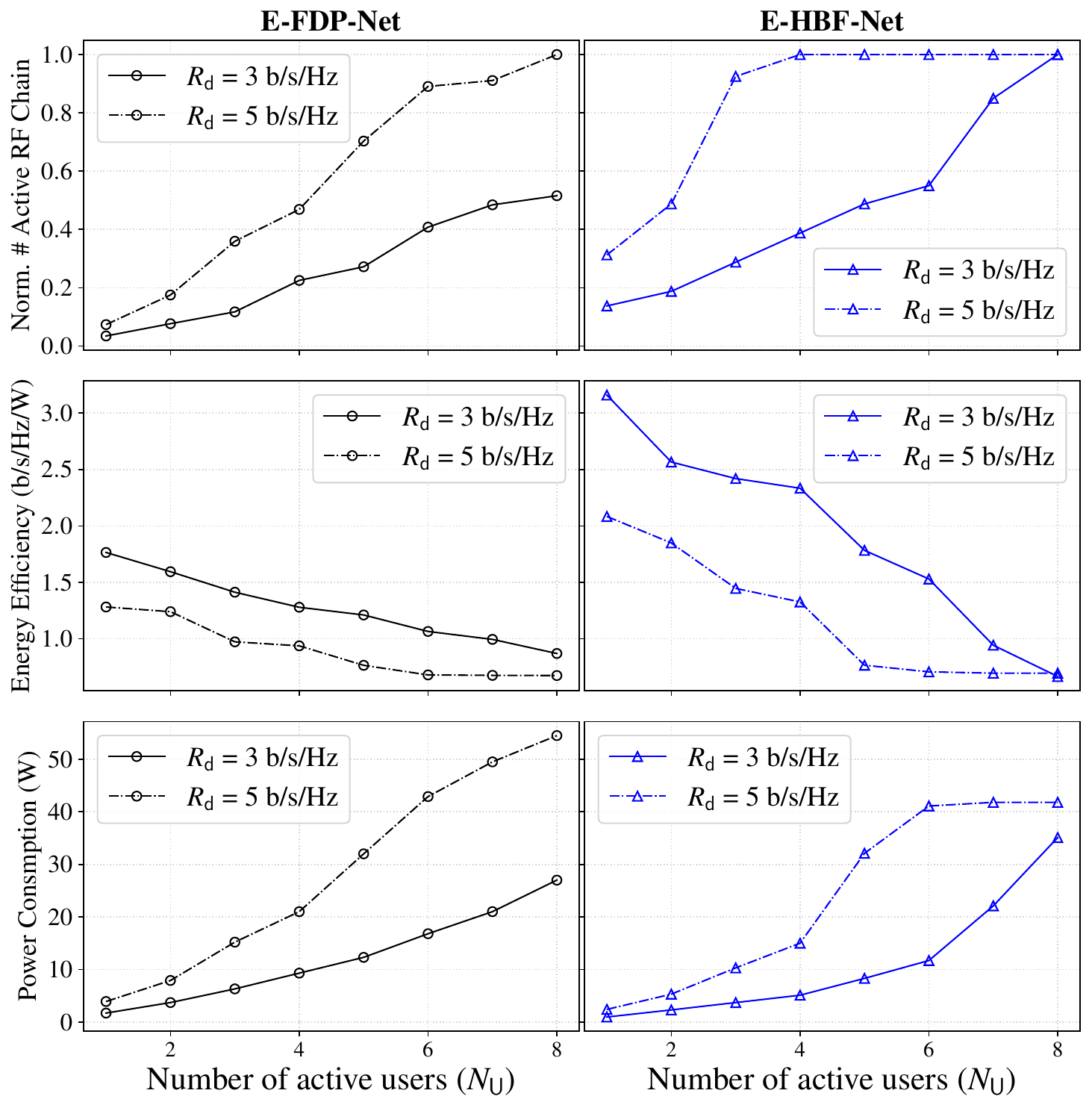}
    \caption{The normalized number of active RF chains, EE, and power consumption of the proposed E-FDP-Net~(left sub-plots), E-HBF-Net~(right sub-plots) versus different numbers of users. System parameters are set to: $N_{\sf{RF}} = 8$, $N_{\sf{T}} = 64$, and $\sigma^2 = -130$ dBm.}
    \label{res:Var_user}
\end{figure}

\subsection{Varying the Number of Users}
To show the impact of antenna and RF chain selection when varying the number of active users, we present Figure~\ref{res:Var_user} for $R_{\sf{d}}=3$ and $R_{\sf{d}}=5$, where the left-side sub-plots present E-FDP-Net and the right-side ones shows E-HBF-Net. To improve the presentation we use the normalized number of active RF chains~($\frac{N_{\sf{RF}}(\bs{\Omega})}{N_{\sf{RF}}}$), which in the case of FDP is equal to the number of active antennas. In the proposed solutions, we see that by increasing the number of active users, the DNN not only activates more RF chains but also increases the transmitted power to meet $R_{\sf{d}}$. Moreover, when $R_{\sf{d}}$ is small, the DNN requires a smaller number of active RF chains while minimizing the transmitted power, thus lowering power consumption and consequently increasing EE. Figure~\ref{res:Var_user} shows that the proposed DNN approaches are adaptive to the number of active users in the network. That is, depending on the scenario, the DNN designs the beamforming structures to adapt to the varying number of users in each scenario. For instance, in a high-traffic scenario, when the number of active users is large, the DNN will activate more antennas and RF chains to meet the average SE. On the other hand, in a low-traffic scenario, when the number of users is low, the DNN has no need to activate a large number of antennas and RF chains, and thus can significantly increase its EE. Finally, we notice that by controlling the value of $R_{\sf{d}}$, which depends on the application and the objective of the service provider, the power consumption can be adjusted.

\subsection{Training with Imperfect CSI}

\begin{figure}[t!]
    \centering
    \includegraphics[width=\columnwidth]{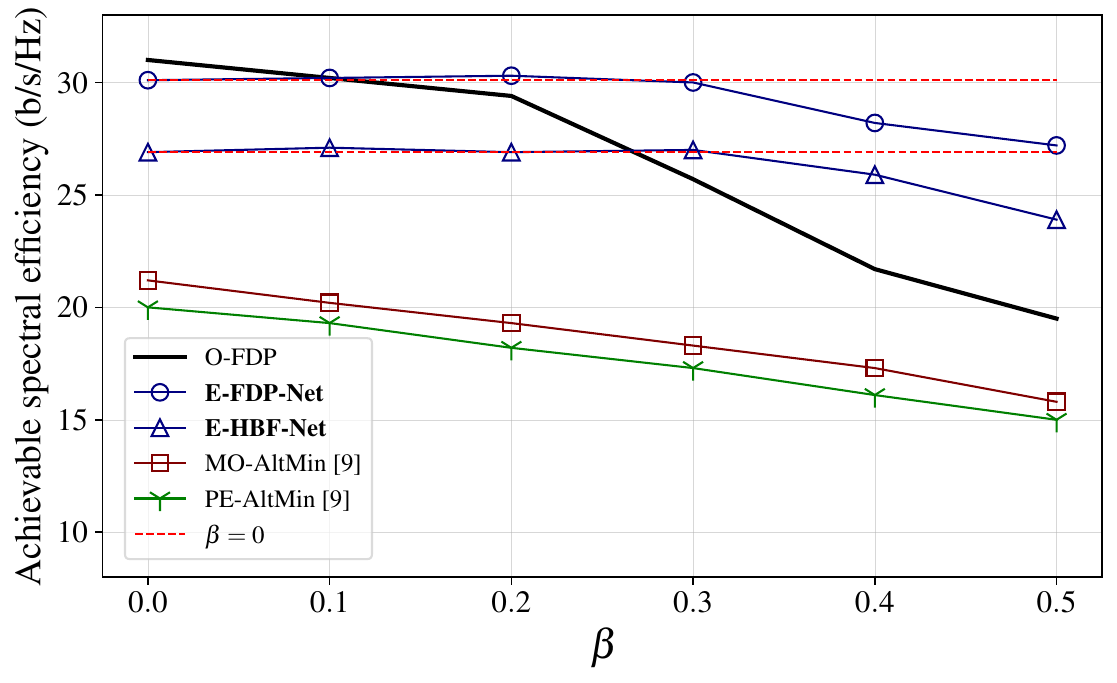}
    \caption{The SE comparison of the different BF solutions with different values of $\beta$. System parameters are set to: $N_{\sf{U}} = 4$, $N_{\sf{RF}} = 8$, $N_{\sf{T}} = 64$, $\gamma = 0$, $\sigma^2=-130$ dBm.}
    \label{res:noisy_csi_all}
\end{figure}

Unlike other studies that assume perfect \gls{CSI} for \gls{DNN} training, in this work, we employed imperfect CSI not only for the input of the \gls{DNN} but also for the computation of the loss function. The robustness of the proposed methods against imperfect CSI is evaluated and compared to other non-DL methods in Figure~\ref{res:noisy_csi_all}. Here we train the \gls{DNN} with different $\beta$ in $\{0, 0.1, 0.2, 0.3, 0.4, 0.5\}$. It is clear that the SE performance decreases as the value of $\beta$ increases. In particular, when $\beta$ increases from $0$ to $0.5$, the SE performance for O-FDP degraded by $38\%$. For PE-AltMin, the degradation is around $25\%$, whereas it is around $27\%$, for MO-AltMin. The lowest degradation in terms of SE performance is achieved for E-HBF-Net and E-FDP-Net, (e.g., the degradation is around $9\%$ and $11\%$, respectively). Therefore, the proposed methods are more robust against estimation errors. Moreover, the red lines in Figure~\ref{res:noisy_csi_all} shows the ideal case of perfect CSI when $\beta =0$. It is interesting to see that for a small $\beta$ (i.e. $0.1$) the SE performance did not degrade, but in contrast, it slightly improved in the online phase. This is due to the fact that training with imperfect CSI can act as a regularization technique known as noise injection in the machine learning literature and thus can improve the generalization of the \gls{DNN} in the online phase~\cite{noiseInje}.


\begin{figure}[t!]
    \centering
    \includegraphics[width=\columnwidth]{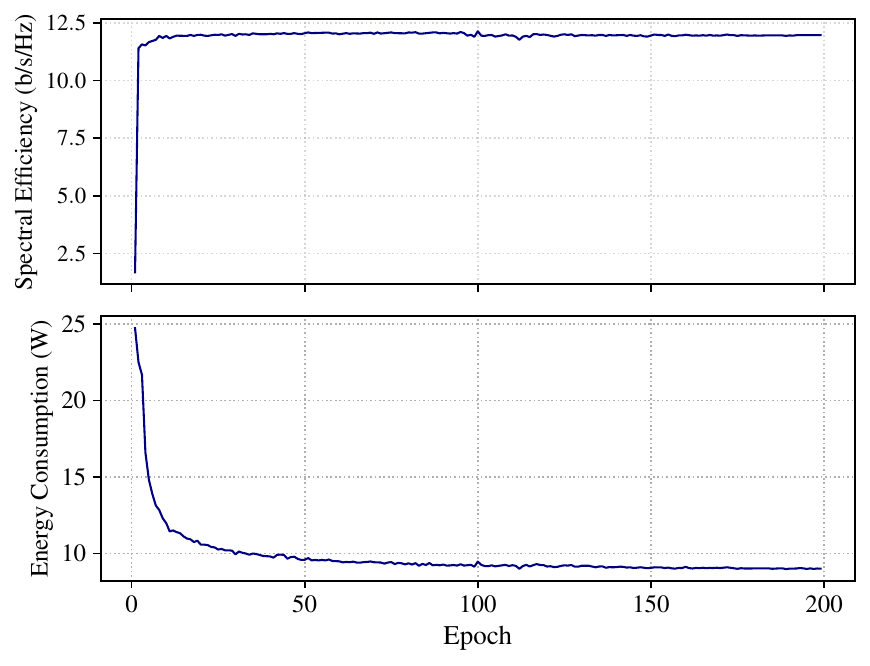}
    \caption{Training steps of E-FDP-Net. The parameters are set to: $N_{\sf{U}} = 4$, $N_{\sf{RF}} = 8$, $N_{\sf{T}} = 64$, and $R_{\sf{d}} = 3$.}
    \label{res:Train}
\end{figure}

In Figure~\ref{res:Train}, we present the convergence of the training of the proposed E-FDP-Net in terms of SE, power consumption, and EE, when $R_{\sf{d}} = 3$ and $N_{\sf{U}}=4$. 
We see in the top subplot that the DNN learns quickly to design the connection matrix and the FDP to obtain an SE of $N_{\sf{U}} R_{\sf{d}} = 12$, i.e., after few epochs, the achieved SE for each user is around $R_{\sf{d}}$. 
Then, while the SE target is respected, the DNN learns to gradually reduce power consumption by turning off some RF chains until it achieves the minimum power consumption as shown in the middle subplot.

\subsection{Computational Complexity Analysis}\label{subsec:comp}
To evaluate the computational complexity of the proposed DNNs, we derive the analytical expression of the number of real multiplications (RM) and compare it with other approaches. We assume that one complex multiplication (CM) corresponds to $4$ RMs and that the $1$ complex division corresponds to $8$ RMs (assuming that the real division of $1$ is equal to $1$ RM). Only the matrix multiplications and inversions are taken into consideration, the other operations are considered negligible. A CM between a matrix of size $N\times P$ and a matrix of size $P\times M$ requires $NMP$ CMs. To invert a square matrix of size $N$, around $N^3/3$ CMs are required if the Gaussian elimination algorithm is employed. Finally, we consider that the eigenvalues of a square matrix of size $N$ are obtained using the Cholesky decomposition~\cite{DD2004}, which requires approximately $4N^3$ RMs.

O-\gls{FDP} requires $4(2^{N_{\sf{U}}} - 1) ( 2 N_{\sf{U}} N_{\sf{T}}^2  +  N_{\sf{U}}^2 N_{\sf{T}}+ \frac{1}{3}N_{\sf{T}}^3)$ RMs as described in~\cite{unsupervised}. In the specified scenario, we replicate the implementation of SoTA algorithms. Our observations reveal that the PE-AltMin algorithm typically achieves convergence within an average of $\ell_{\sf{PE}}=15$ iterations. Given that the computation of the singular-value decomposition of a $p \times q$ matrix necessitates approximately $4p^2 q + 22q^3$ resource modules (RMs), we can formulate the total number of RMs required for PE-AltMin as $\ell_{\sf{PE}} (8N_{\sf{RF}}N_{\sf{U}}(N_{\sf{T}} + N_{\sf{U}}) + 22 N_{\sf{RF}}^3)$.
MO-AltMin has a much higher complexity than PE-AltMin~\cite{yu_tsp_16}. 
MO-AltMin is composed of a main loop that computes the DP, and of an inner loop applying the ``Conjugate Gradient'' algorithm to find the HBF.
In the main loop, computing the DP requires $4N_{\sf{T}}N_{\sf{U}}N_{\sf{RF}}$ RMs, while in the inner loop, the Kronecker product of a $N_{\sf{RF}}\times N_{\sf{T}}$ matrix with a $N_{\sf{U}}\times N_{\sf{T}}$ matrix is computed, which requires $4N_{\sf{T}}^2N_{\sf{U}}N_{\sf{RF}}$ RMs.
Based on the defined scenario the outer loop is repeated $\ell_{\sf{MO}} = 2$ times while the inner loop is repeated $\ell^\prime = 30$ times, the total number of RMs used by MO-AltMin is $4\ell_{\sf{MO}} N_{\sf{T}}N_{\sf{U}} N_{\sf{RF}} \big(1 + \ell^{\prime}N_{\sf{T}} \big)$.
To design the \gls{HBF}, both PE-AltMin and MO-AltMin require designing the FDP as discussed in~\eqref{eq:MMSE_prb}, thus the complexity of obtaining the FDP should be added to the complexity of PE-AltMin and MO-AltMin.

On the other hand, to compute the computational complexity of the DNN approaches, we need to compute the number of parameters of the DNN architectures. Both DNN architectures, E-HBF-Net and E-FDP-Net, have the same $\text{DNN}_{\sf{core}}$ but their output layers are different due to different output dimensions. The number of RMs in the $\text{DNN}_{\sf{core}}$ is calculated for each layer separately, then summed up. The width of the $l^{\text{th}}$ FC and CL are respectively denoted as $f_{l}$  and $c_{l}$. The number of multiplications required for $\text{DNN}_{\sf{core}}$ is $\mathcal{M}(\text{DNN}_{\sf{core}}) = (2c_1 + c_1c_2 + c_2c_3+ c_3f_1/\kappa^2)N_{\sf{T}}N_{\sf{U}}\kappa^2 + f_1f_2$, where $\kappa$ is the kernel size i.e. $\kappa=3$~\cite{unsupervised}. Considering that for E-HBF-Net there are $4$ output layers, one layer for the AP, two layers for the DP, and one layer for the connection matrix, then the total number of multiplications is $\mathcal{M}(\text{DNN}_{\sf{core}}) + f_2(N_{\sf{T}}N_{\sf{RF}} + 2N_{\sf{U}}N_{\sf{RF}} + N_{\sf{T}}N_{\sf{RF}})$. Likewise, for E-FDP-Net, the total number of multiplications is $\mathcal{M}(\text{DNN}_{\sf{core}}) + f_2(2N_{\sf{U}}N_{\sf{T}} + N_{\sf{T}})$. Examples of the numerical values of these analytical expressions are shown in Table~\ref{tbl:comp_co}. 
It can be seen that for HBF transmitters, E-HBF-Net reduces the complexity by $38$\% compared to the least complex conventional approach (PE-AltMin), while for FDP transmitters, O-FDP is $1.5$ times more complex than E-FDP-Net.

\begin{table}[t]
    \centering
    \caption{Computational Complexity Comparison}
    \resizebox{0.8\columnwidth}{!}{
    \begin{tabular}{ccc}
        \toprule
        \multicolumn{1}{c}{Transmitter type} & \multicolumn{1}{c}{Method} & \multicolumn{1}{c}{\# RMs~($\times 10^6$)} \\
        \cmidrule(lr){1-1} \cmidrule(lr){2-2} \cmidrule(lr){3-3}    
        FDP & O-FDP & 7.27 \\
        HBF & PE-AltMin~\cite{yu_tsp_16} & 8.48 \\
        HBF & MO-AltMin~\cite{yu_tsp_16} & 38.7 \\
        FDP & \textbf{E-FDP-Net} & \textbf{4.69} \\
        HBF & \textbf{E-HBF-Net} & \textbf{5.21} \\
        \bottomrule
        \multicolumn{3}{c}{($N_{\sf{U}} = 4$, $N_{\sf{RF}} = 8$ and $N_{\sf{T}} = 64$)}
    \end{tabular}}
    \label{tbl:comp_co}
\end{table}

\section{Conclusion} \label{sec:Conc}
In this paper, we studied the problem of antenna selection and beamforming design in a massive multiple-input multiple-output (mMIMO) system with the objective of maximizing energy efficiency (EE). First, we derived an accurate energy model for the mMIMO system. Our proposed energy model takes into account the transmit power as well as the power consumed by the hardware by considering the insertion loss and the direct power consumption of different components such as the combiners and the power amplifiers. Next, based on our energy model, we designed unsupervised deep learning approaches to intelligently and adaptively select the BF structures and the transmitting antennas. Specifically, we proposed two deep neural networks models, called E-HBF-Net and E-FDP-Net, for hybrid BF and for fully digital precoding, respectively. Both DNNs optimize the EE of the mMIMO system by intelligently selecting the transmitting antennas and choosing the precoding matrices for HBF and FDP, which allows them to achieve significantly better EE than conventional solutions.
Simulation results confirm that the proposed DNNs can adapt to the number of active users and that they provide different trade-offs between SE and EC that can be controlled by tuning a hyper-parameter. 
Furthermore, we show that the DNN models can be trained exclusively using imperfect channel information~(CSI), i.e., the imperfect CSI was used as input to our DNN models as well as to compute the loss function during training. 

\section*{Acknowledgement}
This work was supported by the Natural Sciences and Engineering Research Council of Canada (NSERC) (under project ALLRP 566589-21) and InnovÉÉ (INNOV-R program) through the partnership with Ericsson and ECCC.






\bibliographystyle{IEEEtran}
%
\bibliography{bib/HBF_DL.bib}

\end{document}